\documentclass[onecolumn,showpacs,preprintnumbers,amsmath,amssymb,preprint]{revtex4}

\usepackage{graphicx}
\usepackage[latin1]{inputenc}
\usepackage{dcolumn}
\usepackage{bm}
\usepackage{epsfig}

\begin{document}

\preprint{Preprint}

\title{On the degree distribution of horizontal visibility graphs associated to Markov processes and dynamical systems: diagrammatic and variational approaches}

\author{Lucas Lacasa}
\email{l.lacasa@qmul.ac.uk}
\affiliation{School of Mathematical Sciences\\Queen Mary University of London, Mile End, E14NS London, UK}%

\date{\today}

\begin{abstract}
Dynamical processes can be transformed into graphs through a family of mappings called visibility algorithms, enabling the possibility of (i) making empirical data analysis and signal processing and (ii) characterising classes of dynamical systems and stochastic processes using the tools of graph theory. Recent works show that the degree distribution of these graphs encapsulates much information on the signals variability, and therefore constitutes a fundamental feature for statistical learning purposes. However, exact solutions for the degree distributions are only known in a few cases, such as for uncorrelated random processes. Here we analytically explore these distributions in a list of situations. We present a diagrammatic formalism which computes for all degrees their corresponding probability as a series expansion in a coupling constant which is the number of hidden variables. We offer a constructive solution for general Markovian stochastic processes and deterministic maps. As case tests we focus on Ornstein-Uhlenbeck processes, fully chaotic and quasiperiodic maps.  Whereas only for certain degree probabilities can all diagrams be summed exactly, in the general case we show that the perturbation theory converges. In a second part, we make use of a variational technique to predict the complete degree distribution for special classes of Markovian dynamics with fast-decaying correlations. In every case we compare the theory with numerical experiments.
\end{abstract}

\maketitle

\section{Introduction}
\begin{figure}
\centering
\includegraphics[width=1.\columnwidth]{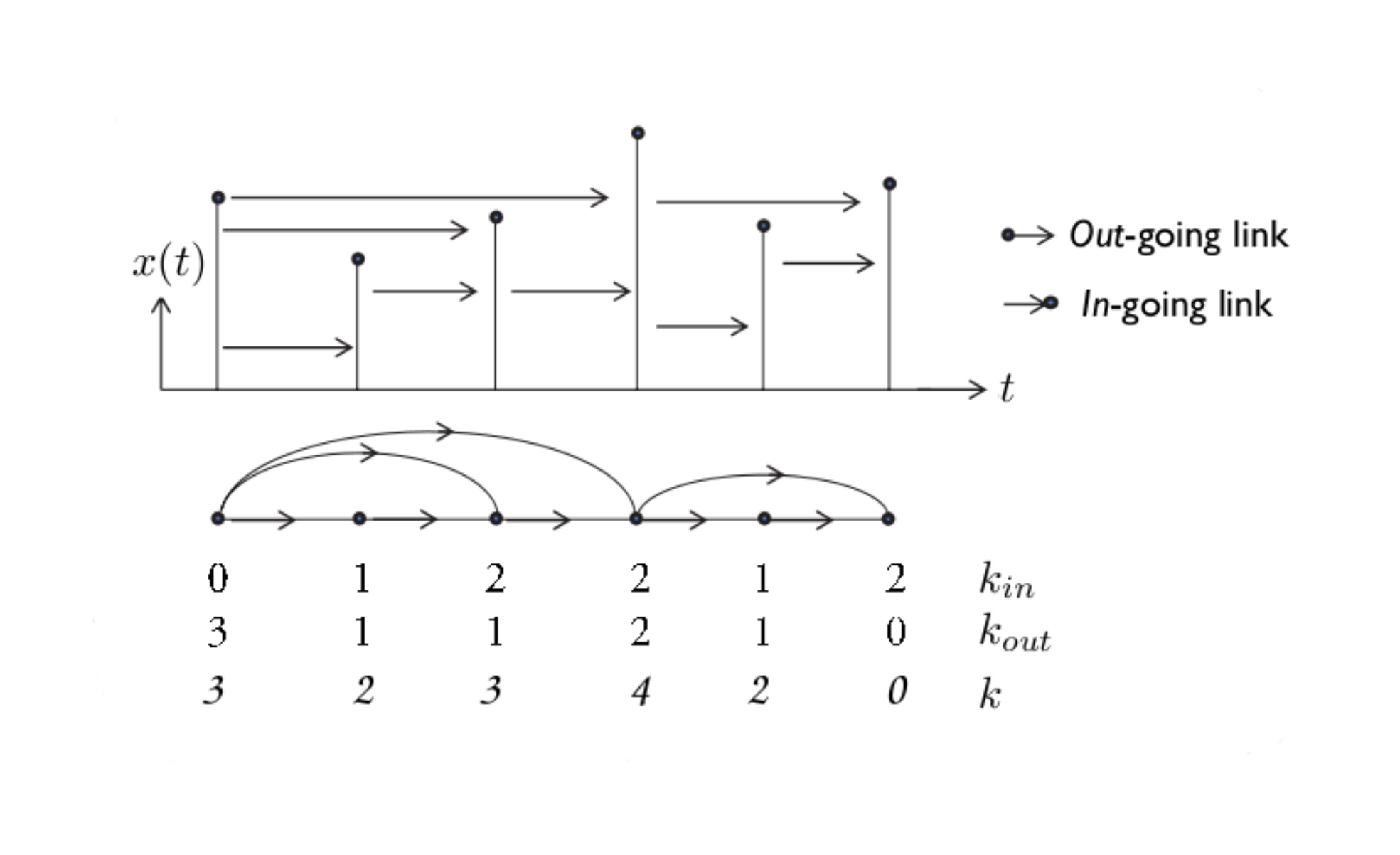}
\vspace{-15mm}
\caption{Graphical Illustration of the Horizontal Visibility Algorithm: in the top we plot a sample time series (vertical bars) and in the bottom we represent its associated Horizontal Visibility Graph (HVG), where each node has a certain degree $k$ and the edges are undirected, and the Directed Horizontal Visibility Graph (DHVG), where each node has an ingoing and an outgoing degree, and the edges are directed. The HVG and DHVG are actually the same graph, although in the DHVG the edges are directed and therefore in this latter case the adjacency matrix is not symmetric. The degree sequence of the HVG and DHVG are also represented.}
\label{fig_label1}
\end{figure}

 The Horizontal visibility algorithm \cite{seminalPRE} is a mapping by which an ordered set of $N$ real numbers $\{x_t\}, t=1,...,N$ maps into a graph $\cal G$ with $N$ nodes and adjacency matrix $A_{ij}$. Nodes $i$ and $j$ are connected through an undirected edge ($A_{ij}=A_{ji}=1$) if $x_i$ and $x_j$ have so called horizontal visibility, that is, if every intermediate datum $x_q$ follows $$x_q<\inf\{x_i,x_j\}, \ \forall q\in [i,j]$$
The set of graphs spanned by this mapping are called Horizontal Visibility Graphs (HVGs). These are noncrossing outerplanar graphs with a Hamiltonian path \cite{Gutin, prodinger}, subgraphs of a more general mapping \cite{pnas} that have been recently used in the context of time series analysis and signal processing \cite{review} (see figure \ref{fig_label1} for an illustration). The methodology proceeds by analysing the topological properties of $\cal G$ and, according to that information, characterise the structure of $\{x_t\}$ and its underlying dynamics. Periodic dynamics are retrieved from the mean degree of the associated graphs, and
some recent applications include the description of correlated stochastic and low-dimensional chaotic series \cite{toral}, processes that seem to cluster as HVGs with different exponential degree distributions $P(k)=a \exp(-\lambda k)$ (we recall that the degree distribution describes the probability of a node chosen at random to have degree $k$). The canonical routes to chaos (Feigenbaum, quasiperiodic, and Pomeau-Manneville scenarios) have also been described in graph theoretical terms \cite{chaos, quasi, intermitencia}, and in particular sensibility to initial conditions has been related, via Pesin identity, to the Shannon and the block entropies of the degree distribution of the associated graphs. Both the horizontal and the original version of the mapping are currently extensively used for data analysis purposes (as feature extraction algorithm for feature-based classification) in several disciplines, such as biomedicine \cite{bio} or geophysics \cite{geo} (see \cite{review} for a recent review).\\

\noindent The fingerprint of the arrow of time and time asymmetries in stationary dynamics can be assessed within this framework by redefining the following node labelling in the transformation: if in the HVG one distinguishes the ingoing degree of a node (where node $i$ has an ingoing edge from $j$ if $j<i$) from its outgoing degree (where node $i$ has an outgoing edge to $j$ if $i<j$), the graph converts into a digraph, so called Directed Horizontal Visibility Graph (DHVG) \cite{irrev}. In a DHVG the ingoing degree distribution $P_\leftarrow(k)$ (the probability that a node chosen at random from $\cal G$ has $k$ ingoing links) and the outgoing degree distribution $P_{\rightarrow}(k)$, defined as the probability that a node chosen at random from $\cal G$ has $k$ outgoing links, are in general different distributions (see figure \ref{fig_label1} for an illustratation). Recent works point that different measures of the distinguishability between $P_{\rightarrow}(k)$ and $P_{\leftarrow}(k)$ \cite{irrev} (amongst other graph properties \cite{irrev2}) can distinguish statistically reversible processes (white noise, stochastic processes with linear correlations, measure preserving chaotic maps) from statistically irreversible stationary processes, such as some dissipative chaotic maps or thermodynamic systems driven out of equilibrium, and are shown to be computationally efficient methods to quantify such asymmetries.\\\
In summary, it appears that the degree distribution of visibility graphs, along with some derived functionals (moments, entropy, etc) carry much information of the signal and the underlying dynamics.\\

\noindent Despite these recent applications, the majority of results using these transformations are numerical and/or heuristic. In this work we focus on the analytical properties of the degree distribution $P(k)$ (and $P_\rightarrow(k)$), when the HVG (or DHVG) is associated to some important classes of dynamical processes. The amount of closed analytical results on these degree distributions is also so far scarce, most of them being found for HVGs which are nontrivial graph theoretical fixed points of some renormalisation group transformation, associated with critical dynamics generated at accumulation points of the canonical routes to chaos \cite{chaos, quasi, intermitencia}. Almost no exact results exist for other nonperiodic dynamical processes, with the exception of uncorrelated random processes \cite{seminalPRE, irrev}. Here we build on previous results and explore in section II how to compute these degree distributions for any given dynamical process with well defined invariant measure. To this end, we propose a general diagrammatic theory similar in spirit to Feynman's approach in quantum theory. Each degree probability $P(k)$ can be computed as a 'perturbative' expansion, where the 'coupling constant' is the number of hidden variables $\alpha$, and for each $\alpha$ a different number of diagrams must me summed up (corrections to the free field). For each $k$, the free field ($\alpha=0$) can be summed up exactly, and diagrams of order $\alpha$ yield a correction of order O($\alpha^{-2}$) in the probability amplitude. In section III we show that the results on uncorrelated processes (for both HVG and DHVG) represent a very special case in this formalism, where all orders of the theory can be summed up exactly, as the $n-$joint distributions completely factorise. We then address in section IV the theory for dynamical processes that fulfill the Markov property and give explicit formulae for a arbitrary diagrams. As case studies we focus on stochastic stationary Markovian processes (Ornstein-Uhlenbeck) in section V and in one dimensional deterministic maps (both chaotic and quasiperiodic) in section VI. For a few terms of the distribution, their diagram expansions can be summed up exactly (up to all orders), but in the general case a convergent perturbative approach should be followed, to calculate each degree probability up to arbitrary precision. In the particular case of chaotic maps, the existence of forbidden patterns \cite{forbidden} drastically decreases the number of allowed diagrams to a finite number, up to a given order, speeding up the convergence of the perturbative analysis.
Then, in section VII we present a general variational approach, by which we can derive analytical results for the entire distribution provided some entropic optimisation hypothesis holds. We show that the entire distributions of chaotic and stochastic Markovian (Ornstein-Uhlenbeck) processes is well approximated if we assume the graphs associated to these processes are maximally entropic. In section VIII we conclude.

\section{A general diagrammatic formalism for degree distributions}
Consider a stationary dynamical process with an underlying invariant density $f(x), \ x \in [a,b]$ (where $a, b \in \mathbb{R}$, and they can be either finite -finite support- or diverge -unbounded support-). $f(x)$ is just a probability density for stationary stochastic processes, or an invariant measure for dynamical systems. Consider also a series of $n$ variables $\{x_0,x_1,...,x_n\}$ extracted from $f(x)$, which can be either a realisation of a stochastic process with underlying density $f(x)$, or a trajectory of some deterministic dynamical system with invariant measure $f(x)$. In both situations, each time series realisation of $n$ variables has a joint probability $f(x_0 \rightsquigarrow x_n)\equiv f(x_0,x_1,x_2,...,x_n)$ (we may call this the propagator). For each realisation of the dynamical process, this ordered set will have an associated HVG/DHVG with a given topology, and in particular with a given degree distribution. As already acknowledged, previous numerical research suggests that such distribution encapsulates and compresses much of the series structure and variability. Consequently, is there any general approach to derive such distributions in a constructive way? The response is positive for large $n$ ($n\rightarrow \infty$), that is, when we consider bi-infinite series and their associated graphs (note, nonetheless, that finite size effects decrease fast with $n$, and therefore whereas theory can only be built in the asymptotic limit, finite size systems converge to the asymptotic solution very fast \cite{seminalPRE}).\\
In what follows we present such a constructive approach. We recall that each datum $x_i$ in the ordered data set is associated with a node with label $i$ in the HVG/DHVG. With a litle abuse of language, from now on we will use $x_i$ to label both the datum and the node (and we will quote as \textit{variable}), although we will make clear when necessary that they represent different mathematical objects.\\

\begin{figure}[h]
\centering
\includegraphics[width=0.50\textwidth]{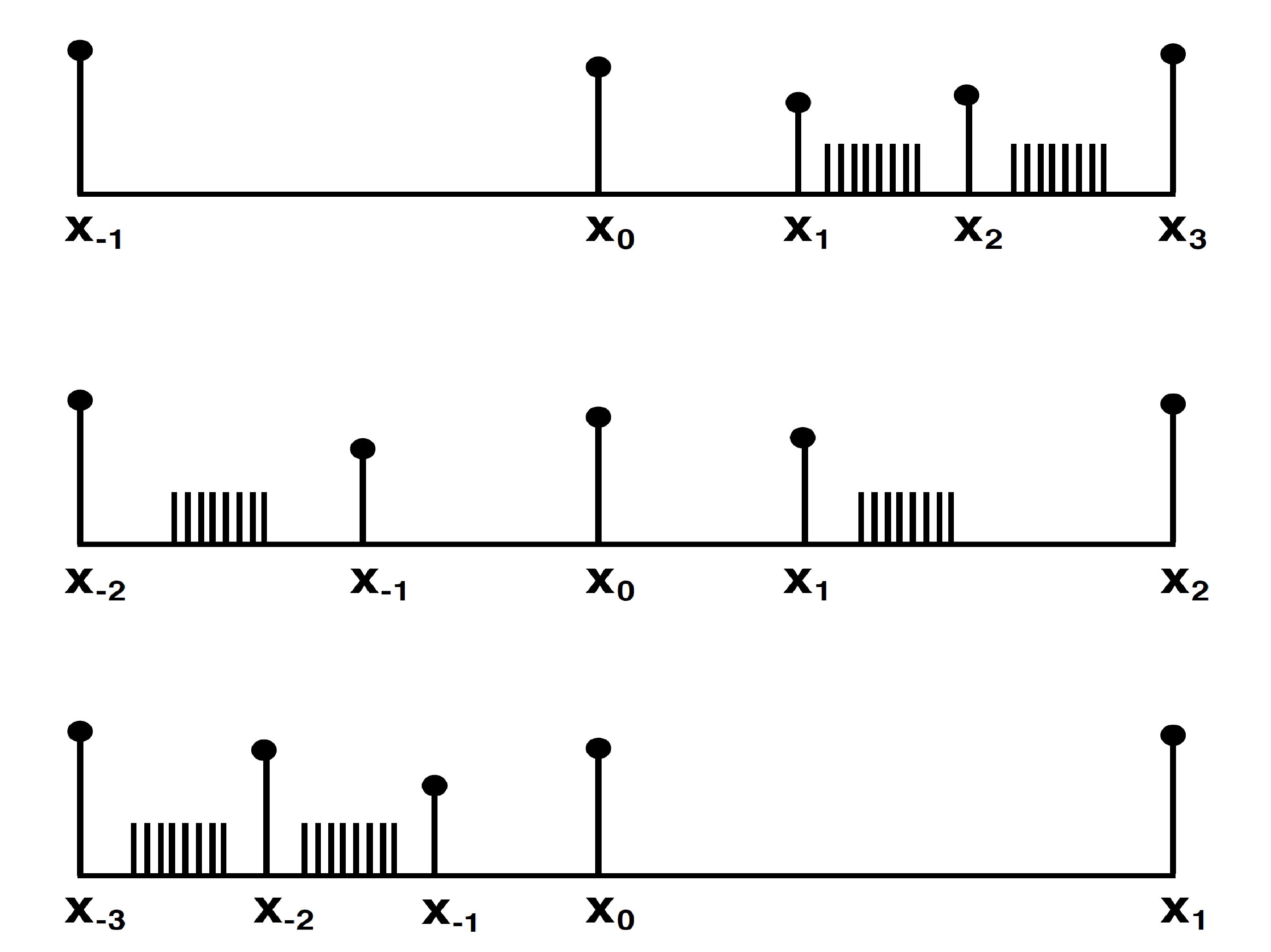}
\caption{Set of possible configurations for a seed variable $x_0$ with
$k=4$. Observe that the sign of the subindex in $x_i$ indicates if
the data is located whether at left-hand side of $x_0$ (sign
minus) or at right-hand side. Accordingly, the bounding's variable
subindex directly indicates the amount of data located in that
side. For instance, $C_0$ is the configuration where none of the
$k-2=2$ inner variables are located in the left-hand side of $x_0$, and
therefore the left bounding variable is labelled as $x_{-1}$ and the
right bounding variable is labelled as $x_3$. $C_1$ is the
configuration for which an inner variable is located in the left-hand
side of $x_0$ and another inner variable is located in its right-hand
side. Finally, $C_2$ is the configuration for which both inner
variables are located in the left-hand side of the seed. Notice that an
arbitrary number of hidden variables can be eventually located among
the inner variables, what is schematically represented in the figure as
a row of vertical lines. }\label{k4}
\end{figure}

Consider a datum (node) chosen at random from the bi-infinite sequence $\{x_t\}$, that we label $x_0$ without loss of generality. To calculate the degree distribution $P(k)$ of the HVG (or the outgoing degree distribution $P_\rightarrow(k)$ of the DHVG) is equivalent to calculate the probability that node $x_0$ has degree $k$ (or outdegree $k$) respectively. For each $k$, a different number of variable configurations (relative positions of data at the right hand side and left hand side of $x_0$) are allowed. Each of these configurations can in turn be schematised as a different \textit{diagram} and will contribute with a different correction to the total probability, as will be shown.\\
For illustrative purposes, consider $P(2)$ of an HVG (as HVGs are undirected and connected, by construction $P(1)=0$). The only configuration that allows $x_0$ to have degree $k=2$ requires, by construction, that the variables on the left and right hand side are larger than $x_0$. Label these \textit{bounding variables} $x_{-1}$ and $x_1$ respectively. Accordingly, this unique configuration (diagram) has an associated contribution to $P(2)$
\begin{equation}
P(2)=\int_{x_0}^ b dx_{-1} \int_{a}^ b dx_{0}\int_{x_0}^ b dx_{1}f(x_{-1},x_0,x_1)
\end{equation}
Incidentally, note at this point that, by construction, the HVGs are outerplanar and have a Hamiltonian path, and therefore $P(2)$ is directly related to the probability that a node chosen at random forms a 3-clique (triplet). Therefore equation \ref{cluster} calculates the global clustering coefficient of a HVG.\\

\noindent A similar calculation can be made for DHVGs. Indeed, $P\rightarrow(0)=0$ (the DHVG is connected) and $P\rightarrow(1)$ is equivalent to calculate the probability that $x_1>x_0$:
\begin{equation}
P\rightarrow(1)=\int_{a}^ b dx_{0}\int_{x_0}^ b dx_{1}f(x_0,x_1)
\end{equation}

 \begin{figure}[h]
\centering
\includegraphics[width=0.8\columnwidth]{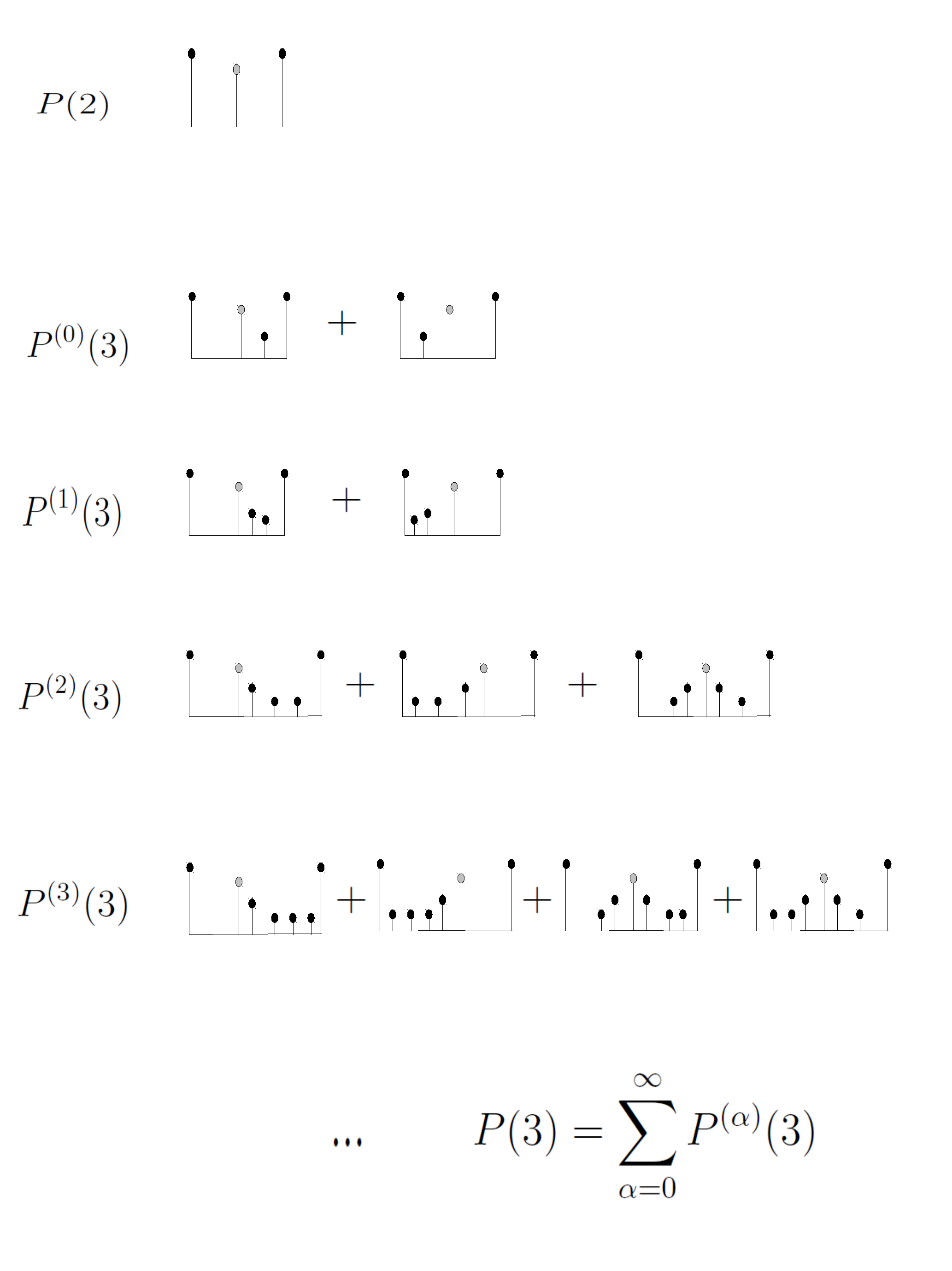}
\caption{Some diagrammatic contributions to the degree distribution $P(k)$ of a horizontal visibility graph (HVG). In each diagram, the grey node corresponds to the reference datum $x_0$, from which the set of concatenated integrals spans (see the text).}
\label{diagrams_HVG}
\end{figure}

\noindent For $P(k\geq 3)$ (or $P\rightarrow(k)\geq 2$) an arbitrary large number of different contributions should be taken into account. Consider first the HVG: if $x_0$ has degree $k\geq3$, then besides the bounding variables, $k-2$ (visible) \textit{inner} variables should be distributed on the right and left hand side of $x_0$. Due to visibility constraints, there are exactly $k-1$ different possible configurations $\{C_i\}_{i=0..k-2}$, where the index $i$
determines the number of inner variables on the left-hand side of $x_0$
 Accordingly, $C_i$ corresponds to the configuration for
which $i$ inner variables are placed at the left-hand side of $x_0$,
and $k-2-i$ inner variables are placed at its right-hand side. Each of
these possible configurations have an associated probability
$p_i\equiv p(C_i)$ that will contribute to $P(k)$ such that
\begin{equation}
P(k)=\sum_{i=0}^{k-2}p_i \label{P_i}.
\end{equation}
In the case of a DHVG, there is only one such configuration $C_0$, such that $P_\rightarrow(k)=p_0 $.\\

\begin{figure}[h]
\centering
\includegraphics[width=0.8\columnwidth]{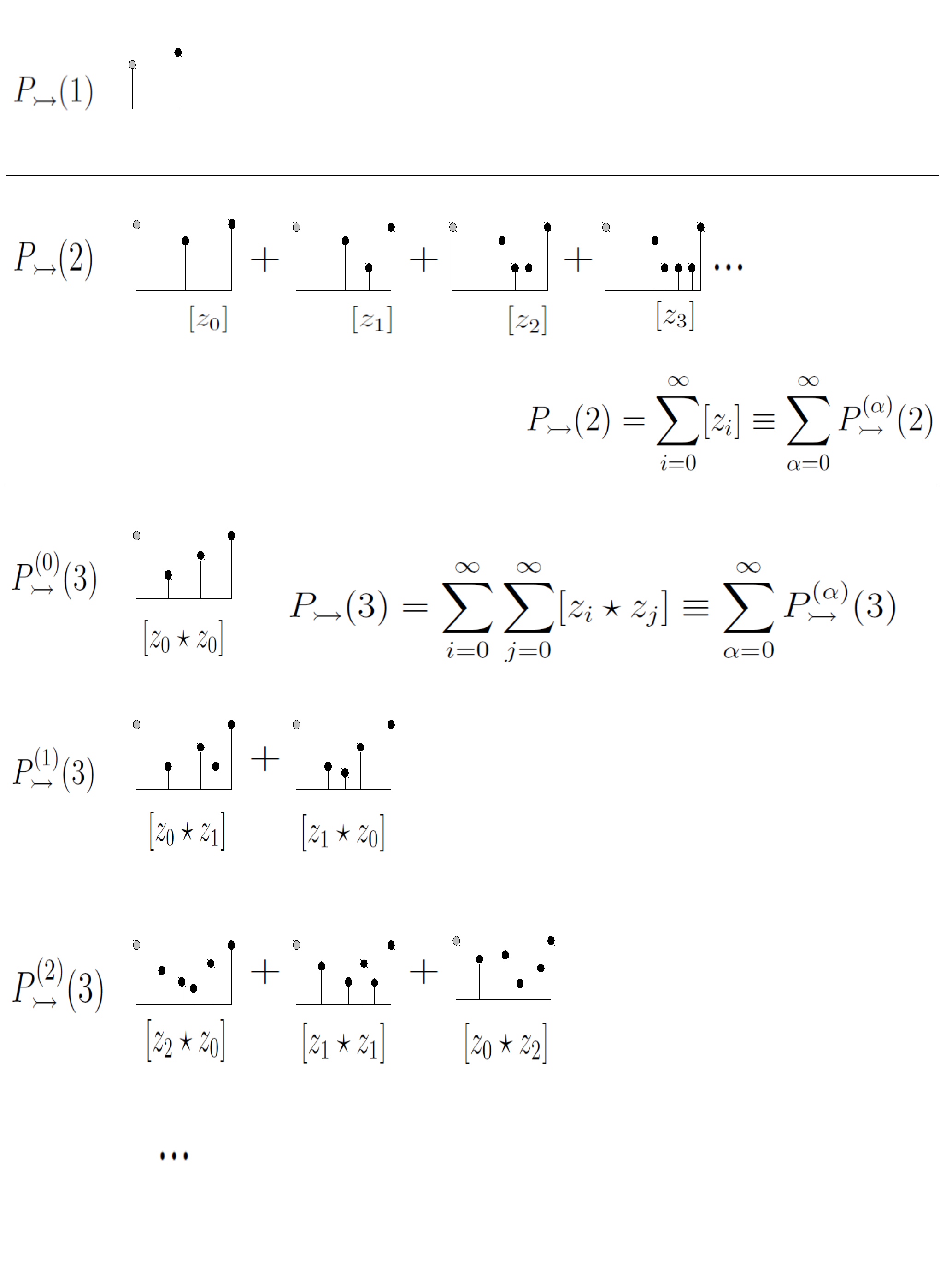}
\caption{Some diagrammatic contributions to the degree distribution $P_\rightarrowtail(k)$ of a directed horizontal visibility graph (DHVG). In each diagram, the grey node corresponds to the reference datum $x_0$, from which the set of concatenated integrals spans (see the text)}
\label{diagramsDHVG}
\end{figure}

\noindent Now, among the inner variables, an arbitrary number of \textit{hidden} variables may appear (see figure \ref{k4} for a graphical illustration of this situation for $P(4)$). In summary, between each pair of inner variables variables $x_i,x_{i+1}$, an arbitrary (eventually infinite) number of hidden variables $\{z_p\}_{p=1}^\infty$ may take place. For each $p$, the different combinations of inner and hidden variables yield different contributions.  As we will show, in some cases it is very convenient to represent such set of contributions as a series expansion 
\begin{equation}
P(k)=\sum_{\alpha} P^{(\alpha)}(k), \label{pertu}
\end{equation}
where $\alpha$ denotes the number of hidden variables. Up to each order, a set of different configurations can take place. Each of these configurations is indeed a different \textit{diagram}. Following Feynman's approach, the free field theory takes into account the contributions spanned by configurations with no hidden variables ($P^{(0)}(k)$), whereas the interaction field introduces corrections of all orders in the number of hidden variables, which is here the coupling constant. Accordingly, $\alpha=0$ accounts for the diagrams with no hidden variables, $\alpha=1$ account for all the diagrams with one hidden variable, etc.\\ 
The same formalism can be extended to DHVGs, where for concreteness we focus on the out degree distribution $P_\rightarrow(k)$
\begin{equation}
P_\rightarrow(k)=\sum_{\alpha} P_\rightarrow^{(\alpha)}(k), \label{pertu2}
\end{equation}
In figures \ref{diagrams_HVG} and \ref{diagramsDHVG} we represent some contributing diagrams up to third order corrections for both HVG and DHVGs.\\

\noindent Two general questions arise:\\
(i) Can we explicitely calculate the contribution of any particular diagram contributing to $P(k)$ or $P_\rightarrow(k)$?\\ 
(ii) What can we say about the convergence properties of the series in (\ref{pertu}, \ref{pertu2})?\\

\noindent Regarding (i), to compute closed form solutions for the entire degree distributions $P(k)$ and $P_\rightarrowtail(k)$ is a hopeless endeavour in the general case, mainly because the n-point propagators  $f(x_0 \rightsquigarrow x_n)$ (where $n$ is also arbitrarily long) cannot be calculated explicitely. However, in several particular cases this propagator factorises and therefore some theory can still be constructed. This is for instance the case of an uncorrelated random process with an underlying probability density $f(x)$. For this large class of stochastic processes, the propagator completely factorises
$$f(x_0,x_1,x_2,...,x_n)=f(x_0)f(x_1)f(x_2)\cdots f(x_n).$$
In the next section we show that this simplification is the key aspect that permits us to calculate $P(k)$ and $P_\rightarrow(k)$ in closed form.

\noindent On relation to (ii), note that all expressions of the form \ref{pertu} or \ref{pertu2} should be converging series as they are probabilities. In particular, a diagram in the correction to the free field of order $\alpha$ has $\alpha$ hidden variables. This diagram is considering the possibility that the seed ($x_0$) and the bounding variable ($x_{k-1}$) are separated by a total of $k-1 + \alpha$ intermediate variables. Now, in a previous work \cite{seminalPRE} it was shown that the probability of two variables separated by $n$ intermediate variables are connected $U(n)$ decreases asymptotically as $n^{-2}$ for uncorrelated random processes. For a constant value of the degree $k$, this means that the correction of the diagrams of order $\alpha$ decreases at least as fast as $\alpha^{-2}$, hence the perturbation series is convergent for the uncorrelated case. For other processes this is not rigorously proved, however we will see that processes with fast decaying correlations are likely to also have associated converging perturbation series.

\section{Uncorrelated processes: exact results for asymptotic distribution}
When the dynamical process under study is a random uncorrelated one, we are in a special case where we don't actually need explicit diagrams to compute the entire degree distribution of both HVG and DHVG, and therefore we don't actually need to follow a perturbative approach such as eqs \ref{pertu} and \ref{pertu2}. We first recall \cite{seminalPRE} the result for HVGs, and further extend this to DHVGs.
 
\subsection{HVG}

\noindent \textbf{Theorem 1.}\\
Let $X(t)$ a real valued bi-infinite time series created from a random variable
$X$ with probability distribution $f(x)$, with $x \in [a,b]$, and
consider its associated Horizontal Visibility Graph $\cal G$. Then, 
\begin{equation}
P(k)=\frac{1}{3}\bigg(\frac{2}{3}\bigg)^{k-2}, \ k=2,3,\dots , \ \forall \ f(x)
\label{uncorr}
\end{equation}

\noindent \textbf{Sketch of the proof.}\\
\noindent The proof proceeds by induction on $k$. Let us begin by computing some easy terms:
\begin{equation}
P(k=2)=\textrm{Prob}(x_{-1},x_1\geq0)
=\int_0^1f(x_0)dx_0\int_{x_0}^1f(x_1)dx_1\int_{x_0}^1f(x_{-1})dx_{-1}.
\label{pk2}
\end{equation}
Now, the cumulative probability distribution function $F(x)$ of
any probability distribution $f(x)$ is defined as
\begin{equation}
F(x)=\int_0^xf(x')dx'\label{cumulative},
\end{equation}
where $dF(x)/dx=f(x)$, $F(0)=0$ and $F(1)=1$. In particular, the
following relation between $f$ and $F$ holds:
\begin{equation}
f(x)F^{n-1}(x)=\frac{1}{n}\frac{dF^n(x)}{dx}.\label{f-F}
\end{equation}
We can accordingly rewrite and compute equation \ref{pk2} as
\begin{equation}
P(k=2)=\int_0^1 f(x_0)[1-F(x_0)]^2dx_0=\frac{1}{3}, \ \forall f(x)
\end{equation}
\noindent In the case $P(k=3)$, two different configurations arise: $C_0$, in which $x_0$
has 2 bounding variables ($x_{-1}$ and $x_2$ respectively) and
a right-hand side inner variable ($x_1$), and the same for $C_1$ but
with the inner variable being place at the left-hand side of the seed:
$$P(k=3)=p(C_0) + p(C_1)\equiv p_0+p_1.$$
\noindent Notice at this point that an arbitrary number $r$ of
hidden variables $n_1,n_2...n_r$ can eventually be located  between the
inner data and the bounding variables, and this fact needs to be taken
into account in the probability calculation. The geometrical
restrictions for the $n_j$ hidden variables are $n_j<x_1, \ j=1,...,r$
for $C_0$ and $m_j<x_{-1}, \ j=1,...,s$ for $C_1$. Then,
\begin{eqnarray}
p_0 &=& \textrm{Prob}\bigg((x_{-1},x_2\geq x_0)\cap (x_1<x_0)\cap (\{n_j<x_1\}_{j=1,\dots,r})\bigg), \nonumber \\
p_1 &=& \textrm{Prob}\bigg((x_{-2},x_1\geq x_0)\cap
(x_{-1}<x_0)\cap(\{m_j<x_{-1}\}_{j=1,\dots,s})\bigg).
\label{P_1_k3bis}
\end{eqnarray}
Now, we need to consider every possible hidden variable configuration. In the particular case of an uncorrelated process,
all these contributions can be summed up exactly:
\begin{eqnarray}
&&p_0= \int_0^1f(x_0)dx_0 \int_{x_0}^1f(x_{-1}) dx_{-1}
\int_{x_0}^1 f(x_2)dx_2
\int_0^{x_0}f(x_1)dx_1 + \nonumber\\
&&\sum_{r=1}^\infty \int_0^1f(x_0)dx_0 \int_{x_0}^1
f(x_{-1})dx_{-1} \int_{x_0}^1 f(x_2)dx_2
\int_0^{x_0}f(x_1)dx_1\prod_{j=1}^r\int_0^{x_1}f(n_j)dn_j\nonumber\label{P0k3}
\end{eqnarray}
where the first term corresponds the contribution of a
configuration with no hidden variables and the second sums up the
contributions of $r$ hidden variables. Making use of the properties of
the cumulative distribution $F(x)$ we arrive to
\begin{equation}
p_0=\int_0^1f(x_0)dx_0 \int_{x_0}^1f(x_{-1}) dx_{-1} \int_{x_0}^1
f(x_2)dx_2 \int_0^{x_0}\frac{f(x_1)}{1-F(x_1)}dx_1,
\end{equation}
where we also have made use of the sum of a geometric series. We
can find an identical result for $p_1$, since the last integral on
equation \ref{P0k3} only depends on $x_0$ and consequently the
configuration provided by $C_1$ is symmetrical to the one provided
by $C_0$. We finally have
\begin{equation}
P(k=3)=2p_0=-2\int_0^1f(x_0)(1-F(x_0))^2\ln
(1-F(x_0))dx_0=\frac{2}{9} \ \forall f(x),
\end{equation}
where the last calculation also involves the change of variable
$z=1-F(x)$.\\

\noindent Hitherto, we can deduce that a given configuration $C_i$
contributes to $P(k)$ with a product of integrals according to the
following rules:
\begin{itemize}
\item The seed variable [S] provides a contribution of
$\int_0^1f(x_0)dx_0$. \item Each boundary variable [B] provides a
contribution of $\int_{x_0}^1 f(x)dx$. \item An inner variable [I]
provides a contribution $\int_{x_j}^{x_0} \frac{f(x)dx}{1-F(x)}$.
\end{itemize}
These 'Feynman rules' allow us to schematize in a formal
way the probability associated to each configuration. For instance
in the case $k=2$, $P(k)$ has a single contribution $p_0$
represented by the formal diagram $[B][S][B]$, while for $k=3$,
$P(k)=p_0+p_1$ where $p_0$'s diagram is $[B][S][I][B]$ and $p_1$'s is
$[B][I][S][B]$. It seems quite straightforward to derive a general
expression for $P(k)$, just by applying the preceding rules for
the contribution of each $C_i$. However, there is still a subtle
point to address that becomes evident for the case
$P(k=4)=p_0+p_1+p_2$ (see figure \ref{k4}). While in this case
$C_1$ leads to essentially the same expression as for both
configurations in $k=3$ (and in this sense one only needs to apply
the preceding rules to derive $p_1$), $C_0$ and $C_2$ are
geometrically different configurations. These latter ones are
configurations formed by a seed, two bounding and two
\emph{concatenated} inner variables, and concatenated variables lead to
concatenated integrals. For instance, applying the same formalism
as for $k=3$, one come to the conclusion that for $k=4$,
\begin{equation}
p_0=\int_0^1 f(x_0)dx_0  \int_0^{x_0} \frac{f(x_1)dx_1}{1-F(x_1)}
\int_{x_1}^{x_0} \frac{f(x_2)dx_2}{1-F(x_2)} \int_{x_0}^1
f(x_3)dx_3 \int_{x_0}^1 f(x_{-1})dx_{-1}.
\end{equation}
While for the case $k=3$ every integral only depended on $x_0$
(and consequently we could integrate independently every term
until reaching the dependence on $x_0$), having two concatenated
inner variables on this configuration generates a dependence on the
integrals and hence on the probabilities. For this reason, each
configuration is not equiprobable in the general case, and thus
will not provide the same contribution to the probability $P(k)$
($k=3$ was an exception for symmetry reasons). In order to weight
appropriately the effect of these concatenated contributions, we
can make use of the definition of $p_i$. Since $P(k)$ is formed by
$k-1$ contributions labelled $C_0, C_1...C_{k-2}$ where the index
denotes the number of inner data present at the left-hand side of
the seed, we deduce that in general the $k-2$ inner variables have the
following effective contribution to $P(k)$:
\begin{itemize}
\item $p_0$ has $k-2$ concatenated integrals (right-hand side of
the seed). \item $p_1$ has $k-3$ concatenated integrals
(right-hand side of the seed) and an independent inner data
contribution (left-hand side of the seed). \item $p_2$ has $k-4$
concatenated integrals (right-hand side of the seed) and another 2
concatenated integrals (left-hand side of the seed). \item ...
\item $p_{k-2}$ has $k-2$ concatenated integrals (left-hand side
of the seed).
\end{itemize}
Observe that $p_i$ is symmetric with respect to the seed.\\

\noindent Including this modification in the Feynman rules,
we are now ready to calculate a general expression for $P(k)$.
Formally,
\begin{equation}
P(k)=\sum_{j=0}^{k-2}[S][B]^2[I]_j[I]_{k-2-j} \label{formal},
\end{equation}
where the sum extends to each of the $k-1$ configurations, the
superindex denotes exponentiation and the subindex denotes
concatenation (this latter expression can be straightforwardly
proved by induction on the number of inner variables). The concatenation of $n$ inner variable integrals
$[I]_n$ reads
\begin{equation}
[I]_n=\int_0^{x_0}
\frac{f(x_1)dx_1}{1-F(x_1)}\prod_{j=1}^{n-1}\int_{x_j}^{x_0}\frac{f(x_{j+1})dx_{j+1}}{1-F(x_{j+1})}.\label{recurrencia}
\end{equation}
which can be proved by induction (using the properties of the cumulative distribution and using appropiate change of variables) to reduce to
\begin{equation}
[I]_n=\frac{(-1)^{n}}{n!}\bigg[\ln
\big(1-F(x_0)\big)\bigg]^n.\label{In}
\end{equation}
According to the formal solution \ref{formal} and to equation
\ref{In}, we finally have
\begin{eqnarray}
P(k)&=&\sum_{j=0}^{k-2}\frac{(-1)^{k-2}}{j!(k-2-j)!}\int_0^1f(x_0)[1-F(x_0)]^2[\ln(1-F(x_0))]^{k-2}dx_0\nonumber
\\
&=& 3^{1-k}\sum_{j=0}^{k-2}\frac{(k-2)!}{j!(k-2-j)!} =
\frac{1}{3}\bigg(\frac{2}{3}\bigg)^{k-2}, \ \forall f(x) \ \ \ \ \square \nonumber 
\end{eqnarray}
\noindent Surprisingly, we can conclude that for every probability
distribution $f(x)$, the degree distribution $P(k)$ of the
associated horizontal visibility graph
has the same exponential form (note that this result parallels the one for the spectrum of white noise, which is universally flat independently of the underlying probability density $f(x)$ of the noise). This is an exact result for sufficiently long uncorrelated random processes (to avoid finite-size or border effects), that is, we consider bi-infinite series ${x(t)}, t\in \mathbb{Z}$ and the calculations address the asymptotic shape of the degree distribution $P(k)$. However, it should be noted that numerical results for finite size series converge fast to the asymptotic solution.

\subsection{DHVG}
\noindent \textbf{Theorem 2.}\\ 
Let $\{x_{t}\}_{t=-\infty,\dots,\infty}$ be a bi-infinite sequence of
independent and identically distributed random variables extracted
from a continuous probability density $f(x)$. Then, the out degree distributions of its associated
directed horizontal visibility graph is
\begin{equation}
P_\rightarrow(k)=\bigg(\frac{1}{2}\bigg)^{k},\ k=1,2,3,\dots , \ \forall f(x)
\label{pk}
\end{equation}
  
\noindent \textbf{Sketch of the proof.}\\
\noindent  The proof follows a similar path as for HVG, such that instead of equation \ref{formal} one gets that for a DHVG and for $k\geq1$
 $$P_\rightarrow(k)=[S][B][I]_{k-1}=\frac{-1^{k-1}}{(k-1)!}\int_0^1f(x_0)(1-F(x_0))\ln [1-F(x_0)]^{k-1}dx_0=\bigg (\frac{1}{2}\bigg)^{k},\  \forall f(x)$$
where we have used the change of variables $z=1-F(x)$ and the formal solution for the concatenation of $n$ inner variable integrals (eq. \ref{In}) $ \ \square \nonumber $ \\

When variables in $\{x_t\}$ are not uncorrelated anymore, the propagator does not factorise and clean, closed form solutions are more difficult to obtain. In what follows we present a general theory for correlated Markovian dynamics. From now on, for concreteness we focus on the DHVG, although a similar formalism  can be extended to HVGs.

\section{Markovian dynamics: a constructive solution}
The second 'easiest' class of dynamical processes are Markovian processes with an integrable invariant measure $f(x)$. For these systems the propagators of the n-joint probabilities factorise into conditional probabilities
$$f(x_0,x_1,x_2,...,x_n)=f(x_0)f(x_1|x_0)f(x_2|x_0)\cdots f(x_n|x_{n-1})$$
Examples include the Ornstein-Uhlenbeck process \cite{kampen} as an example of a stationary stochastic process, the fully chaotic logistic map as an example of an ergodic chaotic map with smooth invariant measure, or irrational rotation as an  example of a zero entropy (nonchaotic), ergodic deterministic process. In what follows we show how to treat this special (although very relevant) class of processes, we develop a formal theory to calculate $P_\rightarrowtail(k)$ in terms of diagrammatic expansions, and we apply it to compute the distributions in the aforementioned examples. \\

For the sake of exposition, let us work out a case by case analysis. For the simplest case $P_\rightarrowtail(1)$, the only possible diagram (see figure \ref{diagramsDHVG}) describes the situations where datum $x_1$ bounds $x_0$ ($x_1>x_0$) and therefore
 \begin{eqnarray}
P_\rightarrowtail(1) = \int_1^ b f(x_0)dx_0 \int_{x_0}^b f(x_1|x_0)
\label{P1}
\end{eqnarray}

\noindent Second, let us consider $P_\rightarrowtail(2)$. In this case we need to take into account the situations where datum $x_0$ sees $x_1$ and $x_2$, that is: $x_2$ is a bounding variable ($x_2\geqslant x_0$) and an arbitrary number of hidden variables can be placed between $x_1$ and $x_2$. There are an infinite number of diagrams that can be labelled as $[z_i]$, where $i$ determines the number of hidden variables in each diagram. Accordingly, 
\begin{eqnarray}
P_\rightarrowtail(2) = \sum_{i=0}^\infty [z_i]\equiv \sum_{\alpha=0}^\infty P_\rightarrowtail^{(\alpha)}(2)
\label{P2}
\end{eqnarray}
where in this case each diagram each correcting order $\alpha$ in the perturbative expansion only includes a single diagram $[z_i]$ that reads
\begin{equation}
[z_0]=\int_{a}^bf(x_0)dx_0 \int_a^{x_0} f(x_1|x_0) dx_1 \int_{x_0}^b f(x_2|x_1)dx_2
\nonumber
\end{equation}
\begin{equation}
[z_1]=\int_{a}^bf(x_0)dx_0 \int_a^{x_0} f(x_1|x_0) dx_1 \int_a^{x_1} f(z_1|x_1)dz_1\int_{x_0}^b f(x_2|z_1)dx_2\nonumber
\end{equation}
\begin{equation}
[z_2]=\int_{a}^bf(x_0)dx_0 \int_a^{x_0} f(x_1|x_0) dx_1 \int_a^{x_1} f(z_1|x_1)dz_1\int_a^{x_1} f(z_2|z_1)dz_2\int_{x_0}^b f(x_2|z_2)dx_2\nonumber
\end{equation}
and for $p>2$,
\begin{equation}
[z_i]=\int_{a}^bf(x_0)dx_0 \int_a^{x_0} f(x_1|x_0) dx_1 \int_a^{x_1} f(z_1|x_1)dz_1 \bigg [\prod_{p=2}^{i-1}\int_a^{x_1} f(z_p|z_{p-1})dz_p\bigg]\int_{x_0}^b f(x_2|z_i)dx_2.
\end{equation}
The third component $P_\rightarrowtail(3)$ is slightly more involved, as there can appear arbitrarily many hidden variables between $x_1$ and $x_2$ and between $x_2$ and $x_3$. Each diagram accounts for a particular combination with $i$ hidden variables between $x_1$ and $x_2$, and $j$ hidden variables between $x_2$ and $x_3$, with the restriction that $x_3$ is a bounding variable and $x_2$ cannot be a bounding variable. There are an infinite number of such contributions, diagrams that can be labelled as $[z_i^{(1)}\star z_j^{(2)}]$:
\begin{equation}
P_\rightarrowtail(3) = \sum_{i=0}^\infty\sum_{j=0}^\infty [z_i^{(1)}\star z_j^{(2)}]\equiv \sum_{\alpha=0}^\infty P_\rightarrowtail^{(\alpha)}(3)
\label{P3}
\end{equation}
where the $[z_i^{(1)}\star z_j^{(2)}]$ diagram, for $i>2$ and $j>2$ yields a correction
\begin{eqnarray}
[z_i^{(1)}\star z_j^{(2)}]&=&\int_{a}^bf(x_0)dx_0 \int_a^{x_0} f(x_1|x_0) dx_1 \int_a^{x_1} f(z_1^{(1)}|x_1)dz_1^{(1)} \bigg [\prod_{p=2}^{i}\int_a^{x_1} f(z_p^{(1)}|z_{p-1}^{(1)})dz_p^{(1)}\bigg]\nonumber \\
&&\ \cdot\int_{x_1}^{x_0} f(x_2|z_i^{(1)})dx_2 \int_a^{x_2} f(z_1^{(2)}|x_2)dz_1^{(2)} \bigg [\prod_{p=2}^{j}\int_a^{x_2} f(z_p^{(2)}|z_{p-1}^{(2)})dz_p^{(2)}\bigg]\int_{x_0}^bf(x_3|z_j^{(2)}) dx_3\nonumber
\end{eqnarray}
As can be seen, in this case more than one diagram contributes to each order in the perturbation expansion. The number of diagrams $N(\alpha, k)$ that contribute to the correction of order $\alpha$ in $P_\rightarrow (k)$ is 
\begin{equation}
N(\alpha, k)= {k-1 \choose \alpha} 
\end{equation}

\noindent Although explicit integral formulae get more and more involved for larger values of the degree $k$, they can be easily expressed diagrammatically. For instance,
\begin{equation}
P_\rightarrowtail(4) = \sum_{i=0}^\infty\sum_{j=0}^\infty\sum_{k=0}^\infty  [z_i^{(1)}\star z_j^{(2)}\star z_k^{(3)}]
\label{P4}  
\end{equation}
describes diagrams with an arbitrary amount of hidden variables located between $x_1$ and $x_2$, $x_2$ and $x_3$, or $x_3$ and $x_4$. The general $[z_i^{(1)}\star z_j^{(2)}\star z_k^{(3)}]$ approximant of this expansion (for $i,j,k>2$) reads:
\begin{eqnarray}
[z_i^{(1)}\star z_j^{(2)}\star z_k^{(3)}]&=&
\int_{a}^bf(x_0)dx_0 \int_a^{x_0} f(x_1|x_0) dx_1 \int_a^{x_1} f(z_1^{(1)}|x_1)dz_1^{(1)} \bigg [\prod_{p=2}^{i}\int_a^{x_1} f(z_p^{(1)}|z_{p-1}^{(1)})dz_p^{(1)}\bigg]\nonumber \\
&&\ \cdot\int_{x_1}^{x_0} f(x_2|z_i^{(1)})dx_2 \int_a^{x_2} f(z_1^{(2)}|x_2)dz_1^{(2)} \bigg [\prod_{p=2}^{j}\int_a^{x_2} f(z_p^{(2)}|z_{p-1}^{(2)})dz_p^{(2)}\bigg] \nonumber \\  
&&\ \cdot\int_{2}^{x_0} f(x_3|z_j^{(2)}) dx_3 \int_a^{x_3}f(z_1^{(3)}|x_3)dz_1^{(3)}\bigg [\prod_{p=2}^{k}\int_a^{x_3} f(z_p^{(3)}|z_{p-1}^{(3)})dz_p^{(3)}\bigg] \nonumber \\  
&&\ \cdot\int_{x_0}^bf(x_4|z_k^{(3)}) dx_4\nonumber
\end{eqnarray}
Finally, by induction we can prove that in the general case the diagrammatic series reads
\begin{equation}
P_\rightarrowtail(k) = \prod_{\Lambda=1}^{k-1} \bigg[\sum_{i{_\Lambda}=0}^\infty\bigg] \bigotimes_{\Lambda=1}^{k-1}[z_{i_\Lambda}^{(\Lambda)}]\bigg (\equiv \sum_{\alpha=0}^{\infty}P_\rightarrow^{(\alpha)}(k)\bigg),
\label{Pk}  
\end{equation}
where the general term $$\bigotimes_{\Lambda=1}^{k-1}[z_{i_\Lambda}^{(\Lambda)}]=
[z_{i_1}^{(1)}\star z_{i_2}^{(2)}\dots \star z_{i_{k-1}}^{(k-1)}]$$
is a diagram that introduces the following correction in order $\alpha=\sum_{j=1}^{k-1}i_j$:
\begin{eqnarray}
\bigotimes_{\Lambda=1}^{k-1}[z_{i_\Lambda}^{(\Lambda)}]&=&
\int_{a}^bf(x_0)dx_0   \prod_{\Lambda=1}^{k-1}\bigg\{
\int_a^{x_0} f(x_1|x_0) dx_1 \int_a^{x_1} f(z_1^{(1)}|x_1)dz_1^{(1)} \bigg [\prod_{p=2}^{i}\int_a^{x_1} f(z_p^{(1)}|z_{p-1}^{(1)})dz_p^{(1)}\bigg]\nonumber \\
&&\ \cdot\int_{x_{i-1}}^{x_0} f(x_i|z_{\Lambda}^{(\Lambda)})dx_i \int_a^{x_i} f(z_1^{(\Lambda)}|x_i)dz_1^{(\Lambda)} \bigg [\prod_{p=2}^{\Lambda}\int_a^{x_i} f(z_p^{(\Lambda)}|z_{p-1}^{(\Lambda)})dz_p^{(\Lambda)}\bigg] \bigg\} \int_{x_0}^b f(x_k|z_{\Lambda}^{(\Lambda)})\nonumber
\end{eqnarray}
Concrete evaluation of each diagram depends on the type of dynamics involved (that is, on the explicit invariant density $f(x)$ and propagator $f(x|y)$), much in the same vein as explicit computation of Feynman diagrams depend on the type of field theory involved \cite{ramond}.\\

\section{Stochastic Markovian dynamics: the Ornstein-Uhlenbeck process}
An Ornstein-Uhlenbeck process \cite{kampen} is a stochastic process that describes the velocity of a massive Brownian particle under the influence of friction. It is the only stationary Gaussian Markov process (up to linear transformations). Since it is a Gaussian process, we have
$$f(x)=\frac{\exp(-x^2/2)}{\sqrt{2 \pi}}$$ 
and the transition probability reads
$$f(x_2|x_1)=\frac{\exp(-(x_2-Kx_1)^2/2(1-K^2))}{\sqrt{2 \pi}(1-K^2)},$$
where $K=\exp(-1/\tau)$ and the correlation function is $C(t)\sim \exp(-t/\tau)$. Variables are Gaussian, therefore $a=-\infty$ and $b=\infty$. Note that in this case we also expect convergent perturbation expansions of the type (\ref{pertu2}), since calculation of long distance visibility $U(n)$ for short-range correlated processes can be approximated, for large $n$, to the uncorrelated case.

\subsection{Exact solutions}
\noindent To begin, consider $P_\rightarrowtail(1)$. Note that this probability is only spanned by a single diagram (no hidden variables), yielding simply
$$P_\rightarrowtail(1)=\int_{-\infty}^{\infty}dx_0f(x_0)\int_{x_0}^{\infty} dx_1f(x_1|x_0)=\frac{1}{2},$$
for $\tau=1.0$.\\

\noindent $P_\rightarrowtail(2)$ is more involved, as it has corrections to the free field solution at all orders (see figure \ref{diagramsDHVG}). In the next section we compute a few terms of this diagrammatic expansion. Here we show that a mathematical trick can be used to compute all orders of $P_\rightarrowtail(2)$ simultaneously, yielding an exact result. The technique is numerical but nonetheless quite sound. We start by rewriting the contribution of each diagram in equation \ref{P2} as a recursive relation
\begin{eqnarray}
P_\rightarrowtail(2)=\int_{-\infty}^{\infty} f(x_0) \sum_{p=0}^\infty I(p|x_0)\nonumber
\label{P+2OU}
\end{eqnarray}
where $I(p|x_0)$ satisfies
\begin{equation}
I(p|x_0)=\int_{-\infty}^{x_0} dx_1f(x_1|x_0) G_p(x_1,x_1,x_0),\label{generalI} 
\end{equation}
and $G_p$ satisfies:
\begin{eqnarray}
G_0(x,y,z)&\equiv&\int_z^\infty f(h|y)dh, \\
G_p(x,y,z)&=&\int_{-\infty}^x dh f(h|y)G_{p-1}(x,h,z), \ p\geq1. \label{Ggen}
\end{eqnarray}
(These latter recursions can be straightforwardly proved by induction on the number of hidden variables). The last equations represent a convolution that can be one more time formally rewritten as $G_p=TG_{p-1}$, or $G_p=T^pG_0$,
with an integral operator $T=\int_{-\infty}^x dh f(h|y)$. Accordingly,

\begin{equation}
I(p|x_0)=\int_{-\infty}^{x_0} dx_1f(x_1|x_0) \sum_{p=0}^{\infty}G_p(x_1,x_1,x_0)\equiv\int_{-\infty}^{x_0} dx_1f(x_1|x_0) S(x_1,x_1,x_0).
\label{p2m}
\end{equation}
In the last equation the formal sum $S(x,y,z)$ is defined as
\begin{equation}
 S(x,y,z)=\sum_{p=0}^\infty G_p(x,y,z)=\sum_{p=0}^\infty T^p G_0=\frac{1}{1-T}G_0,\label{S1}
\end{equation}
where the convergence of last summation is guaranteed provided the spectral radius $r(T)<1$, that is,
\begin{equation}
\lim_{n\rightarrow\infty} ||T^n||^{1/n}<1,
\end{equation}
where $||T||=\max_{y\in(-\infty,x)} \int_{-\infty}^x dh |f(h|y)|$ is the norm of $T$. This condition is trivially fulfilled as $f(x|y)$ is a
Markov transition probability. Then, equation (\ref{S1}) can be written as $(1-T)S=G_0$, or more concretely
\begin{equation}
S(x,y,z)={G}_0(x,y,z) + \int_{-\infty}^x dh f(h|y) S(x,h,z), \label{SSS}
\end{equation}
which is a Volterra equation of the second kind for $S(x,y,z)$. 
Typical one-dimensional Volterra integral equations can be numerically solved applying quadrature formulae for approximate the integral operator \cite{librodeernesto}. The technique can be easily extended in the case that the integral equation involves more than one variable, as it is this case, using a Simpson-type integration scheme to compute the function $S(x,y,z)$. One technical point is that one needs to replace the $-\infty$ limit in the integral by a sufficienly small number $a$. We have found that $a=-10$ is enough for a good convergence of the algorithm. Given a value of $z$ the recursion relation  
\begin{eqnarray}
&&S(a,a+n\delta,z)= G_0(a,a+n\delta,z)\nonumber \\
&&S(a+k\delta,a+n\delta,z)= G_0(a,a+n\delta,z)+\delta\sum_{i=0}^{k-1}f(a+i\delta|a+n\delta)S(a+(k-1)\delta,a+i\delta,z)+ O(\delta^2)\nonumber \label{SS}
\end{eqnarray}
where $\delta$ is the step of the algorithm, for $k=0,1,2,\dots$ and $n=0,1,\dots,k$, allows us to compute $S(x,y,z)$ for $y\le x$. We find (for a correlation time $\tau=1.0$)
$$P_\rightarrowtail(2)\approx 0.24,$$
on good agreement with numerical experiments (table \ref{table2}).\\

\subsection{Perturbative expansions}
\noindent In general the infinite set of diagrams that contributes to each probability $P_\rightarrowtail(k)$ cannot be summed up in closed form, and a perturbative approach should be followed. Here we explore the accuracy and convergence of such series expansion (eq. \ref{pertu2}).\\

\noindent $\underline{\alpha=0}.$\\
\noindent Up to zeroth order (free field theory), we count only diagrams with no hidden variables. Accordingly,
the first degree probabilities are
\begin{eqnarray}
P_\rightarrowtail^{(0)}(1)=\int_{-\infty}^{\infty}dx_0f(x_0) \int_{x_0}^{\infty} dx_1f(x_1|x_0)\nonumber \\
P_\rightarrowtail^{(0)}(2)=\int_{-\infty}^{\infty}dx_0f(x_0)\int_{-\infty}^{x_0} dx_1f(x_1|x_0)
\int_{x_0}^{\infty} dx_2f(x_2|x_1)\nonumber \\
P_\rightarrowtail^{(0)}(3)=\int_{-\infty}^{\infty}dx_0f(x_0)\int_{-\infty}^{x_0} dx_1f(x_1|x_0)
\int_{x_1}^{x_0} dx_2f(x_2|x_1)\int_{x_0}^{\infty}dx_3f(x_3|x_2)\nonumber \\
\cdots
\end{eqnarray}
These can be calculated up to arbitrary precision. As an example, for a correlation time $\tau=1.0$ and approximating $(-\infty,\infty)$ by $(-4,4)$ (note that this is a good approximation as the OU process is Gaussian), we find
\begin{eqnarray}
P_\rightarrowtail^{(0)}(1)=0.500\nonumber \\
P_\rightarrowtail^{(0)}(2)=0.151\nonumber \\
P_\rightarrowtail^{(0)}(3)=0.043 \nonumber
\end{eqnarray}
Now, $P_\rightarrowtail^{(0)}(1)$ is exact as long as $P_\rightarrowtail(1)$ does not admit hidden variables.\\\\
 
\noindent $\underline{\alpha=1}.$\\
\noindent The interaction field couples an arbitrary number of hidden variables, distributed amongst the $k-1$ inner variables. Up to first order (figure \ref{diagramsDHVG}), we further count diagrams with one hidden variable. For instance,
\begin{eqnarray}
P_\rightarrowtail^{(1)}(2)=\int_{-\infty}^{\infty}dx_0f(x_0)\int_{-\infty}^{x_0} dx_1f(x_1|x_0)
\int_{-\infty}^{x_1}f(z_1|x_1)dz_1\int_{x_0}^{\infty} dx_2f(x_2|z_1)\approx 0.0366\nonumber 
\end{eqnarray}
Therefore, up to first order, the analytical predictions are
\begin{eqnarray}
P_\rightarrowtail^{\text{th}}(1)=0.5\nonumber \\
P_\rightarrowtail^{\text{th}}(2)=0.19\nonumber 
\end{eqnarray}
Similar calculations can be performed for other values of the degree $k$ and for higher order corrections $\alpha \geq 2$, in order to reach arbitrarily accurate estimations. The only restriction we have in this example is the power of our symbolic calculator, as we need to perform symbolic concatenated integrals of Gaussian functions.
\begin{table}
\begin{ruledtabular}
\begin{tabular}{ccccc}
degree $k$ &$P_\rightarrowtail(k)$ (numerics)&$P_\rightarrowtail(k)$ (exact)&$P_\rightarrowtail^{(0)}(k)$  &$P_\rightarrowtail^{(0)}(k)+P_\rightarrowtail^{(1)}(k)$\\
\hline
1&0.50&0.5&0.5&0.5\\
2&0.24&0.24&0.151&0.19\\
3& 0.12&?&0.043&\\
\end{tabular}
\end{ruledtabular}
\caption{\label{table2} Comparison of numerical results and theoretical predictions for the out degree distribution of DHVG associated to an Ornstein-Uhlenbeck process (see the text) with $\tau=1.0$. Numerics are the result of simulations, where a trajectory of $2^{20}$ time steps extracted from the map is mapped to its DHVG and the degree distribution is numerically computed. For $k=1$ perturbation theory at zeroth order is exact due to the restriction of forbidden patterns (see the text). For $k=2$ we can see that up to first order we already reach a reasonably accurate result.}
\end{table}

\noindent To evaluate the accuracy of zeroth and first order in the general case, we make some numerical simulations. We generate time series of $2^{20}$ data extracted from an OU process with the same characteristics, transform the series into a DHVg and compute its degree distribution. A comparison with our theory is shown in table \ref{table2}.
The first term ($k=1$) is exact in our theory and therefore coincides with the numerical experiments. The results of zeroth and first order corrections for $k=2,3$ are still far away from the experiments, however convergence guarantees that such methodology provides an arbitrary accurate result.\\ 
In the following section we address deterministic maps and show that convergence of the perturbative approach is faster than for stochastic processes.

\section{One dimensional deterministic maps with smooth invariant measure - the chaotic and quasiperiodic cases}

A deterministic map of the form $x_{t+1}=H(x_t)$ also has the Markov property, and therefore the same mathematical framework applies, provided that there exists an (invariant) probability measure $f(x)$ that characterizes the long-run proportion of time spent by the system in the various regions of the phase space. In the case of dissipative chaos, $f(x)$ describes the distribution of visits to different parts of the map's attractor. If such attractor has integer dimension, then the invariant measure is at least some piecewise continuous function and therefore integrable: in that case the general methodology presented in section IV holds. For chaotic maps with fractal attractor, a more general integration theory should be adopted (see the discussion section).\\
As $H$ is a deterministic map, the transition probability $f(x|y)$ (the propagator) reads
$$f(x|y)=\delta(x-H(y)),$$
where $\delta(x)$ is the Dirac generalised function. This means that nested integrals that appear in each diagram only yield a binary ($0/1$) result.
As we will see, this implies that the only effect of these nested integrals is to rescale the range of integration of $x_0$. For instance, in the case of dissipative chaotic maps this rescaling is in turn associated to the fixed point structure of the $n$th iterates $\{H(x), H^{(2)}(x),\cdots,H^{(n)}(x)\}$ (for piecewise linear maps, other criteria should be followed). Let us illustrate this with the calculation of $P_\rightarrowtail(1)$ associated to
a general one dimensional deterministic map $x_{t+1}=H(x_t)$. Recall that $P_\rightarrowtail(1)$ does not allow hidden variables and therefore its computation is only related to a single diagram, whose probability is
$$P_\rightarrowtail(1)=\int_a^b f(x_0)dx_0\int_{x_0}^b \delta (x_1-H(x_0))dx_1.$$
Now, interestingly, the dependent integral vanishes $\forall x_0$ if $H(x_0)<x_0$, and is equal to one otherwise. Therefore, this rescales the integration range of $x_0$ 

$$[a,b]\Longrightarrow [a',b'],$$
such that for $x_0 \in [a',b']$, $H(x_0)<x_0$. Hence and
$$P_\rightarrowtail(1)=\int_{a'}^{b'} f(x_0)dx_0$$
If the map is unimodal, then $a'=x^*$ and $b'=b$, where $x^*$ is the unstable fixed point of $H$ satisfying $H(x^*)=x^*$. In general, in order to compute each of the diagrams associated to $P_\rightarrowtail(k)$, the $n$-th nested integral rescales the integration range according to the fixed points of $H^{n}(x)$. This property makes the calculation of these diagrams in the case of chaotic dynamics much easier than for stochastic Markovian dynamics. Moreover, the fact that the chaotic trajectories visit the phase space in an orchestrated way introduces forbidden patterns \cite{forbidden} in the temporal order of visits to regions of the phase space. This further introduces 'forbidden diagrams', what will reduce significantly the number of corrections that need to be counted up to each other $\alpha$. On top of that, note that in this case we also expect convergent perturbation expansions of the type (\ref{pertu2}): as the correlation function $<x_t x_{t+\tau}$ of chaotic processes vanishes for distant data ($\tau<<1$), long distance visibility $U(n)$ for chaotic processes can be approximated, for large $n$, to the uncorrelated case. In what follows we focus on two concrete maps: a chaotic map with integer dimension (the fully chaotic logistic map) and a nonchaotic quasiperiodic map.

\subsection{Chaotic dynamics}
\begin{figure}
\centering
\includegraphics[width=0.9\columnwidth]{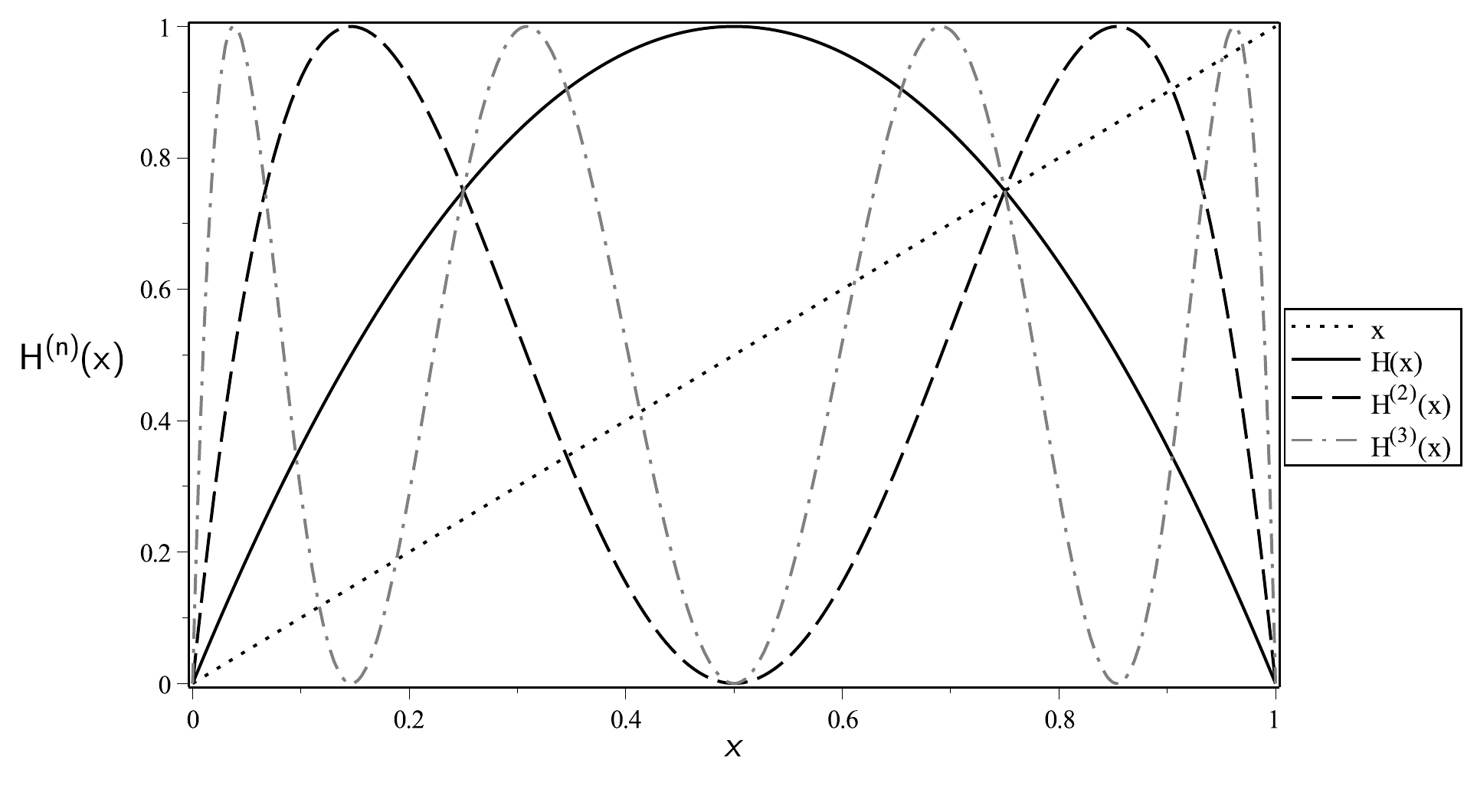}
\caption{Cobweb plot of the fully chaotic logistic map $H(x)=4x(1-x)$ and some of its iterates $H^{(n)}(x)$. The specific intervals where $H^{(n)}(x)>(<)x$ depend on the distribution of fixed points $\{x^*\}$, that fulfill $H^{(n)}(x^*)=x^*$.}
\label{lm}
\end{figure}

For the sake of concreteness, we focus on logistic map $H(x_n)=\mu x_n(1-x_n)$ with parameter  $\mu = 4$, where the map is ergodic, the attractor is the whole
interval $[0,1]$ and the invariant measure $f(x)$ corresponds to a Beta distribution 
\begin{equation}
f(x)=\frac{1}{\pi \sqrt{x(1-x)}} 
\label{rho}
\end{equation}
This is topologically conjugate to the tent and Bernoulli maps. Its fixed points and unstable periodic orbits (fixed points of $H^{(n)}$) are 
\begin{eqnarray}
&& H^{(1)}(x)=4x(1-x), \ x^*\equiv S_1=\{0,3/4\}\nonumber \\
&& H^{(2)}(x)=-16x(x-1)(4x^2-4x+1), \ x^*\equiv S_2=\{0, 3/4, \frac{5-\sqrt{5}}{8}, \frac{5+\sqrt{5}}{8}\}\nonumber \\
&& H^{(3)}(x)=-64x(x-1)(4x^2-4x+1)(64x^4-128x^3+80x^2-16x+1), \ x^*\equiv  S_3 \label{fp}
\end{eqnarray}
where $S_n, \ n\geq3$ can only be computed numerically according to Abel-Ruffini's theorem (see figure \ref{lm} for a cobweb diagram of several iterates $H^{(n)}(x)$).
In what follows we study the diagrammatic expansion for all degree $k$.
To begin, consider $P_\rightarrowtail(1)$, generated by a single diagram
\begin{equation}
P_\rightarrowtail(1)=\int_0^1 f(x_0)dx_0\int_{x_0}^1 \delta (x_1-H(x_0))dx_1.
\end{equation}
The second integral is one for $H(x_0)>x_0$ (and zero otherwise), that is, for $x_0<3/4$, where $x^*=S_1=\{0,3/4\}$ are the fixed points of $H(x)$. Therefore,
\begin{equation}
P_\rightarrowtail(1)=\int_0^{3/4} \frac{dx_0}{\pi\sqrt{x(1-x)}}=\frac{2}{3}
\end{equation}

\noindent In a second step, consider $P_\rightarrowtail(2)$, which is generated by an infinite number of diagrams, each of which with $n$ hidden variables. This again can be written down perturbatively

$$P_\rightarrowtail(2) = \sum_{\alpha=0}^{\infty}P_\rightarrowtail^{(\alpha)}(2).$$ At zeroth order, the diagram reads
\begin{equation}
P_\rightarrowtail^{(0)}(2)= \int_0^1 f(x_0)dx_0\int_{0}^{x_0} \delta (x_1-H(x_0))dx_1\int_{x_0}^1 \delta(x_2-H(x_1))dx_2.
\end{equation}
The last integrals rescale the integration range of $x_0$ according to the solutions of:
\begin{eqnarray}
H(x_0)<x_0\nonumber\\
H^{2}(x_0)>x_0.
\label{GEN1}
\end{eqnarray}
The first condition rescales $x_0 \in [3/4,1]$. For the second condition, we use the fixed points set $S_2$ (see  (\ref{fp})), and after a little algebra we find that if we label $p=\frac{5-\sqrt{5}}{8}; q=\frac{5+\sqrt{5}}{8}$, we get 
$$H^{2}(x_0)>x_0 \Longleftrightarrow x_0 \in [0,p] \cup [3/4,q].$$
The intersection of both conditions is then $[3/4,(5+\sqrt{5})/8$, hence
\begin{equation}
P_\rightarrowtail^{(0)}(2)= \int_{3/4}^{\frac{5+\sqrt{5}}{8}} \frac{dx_0}{\pi\sqrt{x(1-x)}}=\frac{2}{15}
\end{equation}
Now, the rest of orders require $\alpha>0$ hidden variables. In every such diagram (see figure \ref{diagramsDHVG}), an implicit condition is $x_1<x_0$ and $z_1<x_0$, where $z_1$ is the first hidden variable. However, this is a forbidden 3-pattern: there are no three consecutive points in any orbit of the logistic map forming a strictly decreasing trio \cite{forbidden} Therefore we conclude that the infinite number of diagrams with hidden variables do not further contribute to $P_\rightarrowtail(2)$, and in this special case the probability is exactly solvable
\begin{eqnarray}
&&P_\rightarrow^{(0)}(2)=2/15 \nonumber \\
&&P_\rightarrow^{(\alpha >0 )}(2)=0\nonumber \\
&&\Rightarrow P_\rightarrow(2)=\frac{2}{15}
\end{eqnarray}

\noindent Let us proceed forward with $P_\rightarrowtail(3)$. We will show that the diagrams associated to each perturbation order $\alpha$ can be formally calculated using the fixed point structure of $H^{(n)}(x)$.\\
 
\noindent $\underline{\alpha=0}$\\
The zeroth order diagram (no hidden variables) contributes with only one diagram such that
$$P_\rightarrowtail^{(0)}(3)=\underbrace{\int_0^1 f(x_0)dx_0}_{I_0}\underbrace{\int_0^{x_0} \delta (x_1-H(x_0))dx_1}_{I_1} \underbrace{\int_{x_1}^{x_0} \delta(x_2-H(x_1))dx_2}_{I_2} \underbrace{\int_{x_0}^1 \delta(x_3-H(x_2))dx_3}_{I_3}.$$ 
Each integral $I_m$ rescales the integration range in $I_0$ according to the intersection of the solution interval of the following inequalities

\begin{eqnarray}
I_1\rightarrow H(x_0)<x_0\nonumber\\
I_2\rightarrow H^{2}(x_0)<x_0\nonumber\\
I_3\rightarrow H^{3}(x_0)>x_0
\label{GEN2}
\end{eqnarray}
We have:
$$H(x_0)<x_0\Rightarrow x_0 \in [3/4,1].$$
On top of that, the second inequality reduces this range further:
$$H(x_0)<x_0 \cap H^{2}(x_0)<x_0 \Rightarrow x_0 \in [\frac{5+\sqrt{5}}{8},1].$$
Last inequality cannot be solved in an algebraic closed form (the order of the polynomial is larger than five), but the solution can be calculated numerically up to arbitrary precision, finding that the intersection interval is
$$H(x_0)<x_0 \cap H^{2}(x_0)<x_0 \cap H^{3}(x_0)>x_0\Rightarrow x_0 \in [.95, .97],$$
therefore
$$P_\rightarrowtail^{(0)}(3)\approx \int_{0.95}^{0.97} \frac{dx_0}{\pi \sqrt{x(1-x)}}=0.0315$$

\noindent $\underline{\alpha=1}$\\
Note that for $\alpha=1$ (diagrams with one hidden variable), only one of the two possible diagrams is allowed, as the other one is again a forbidden one. 
The allowed diagram yields a correction
\begin{eqnarray}
&&P_\rightarrowtail^{(1)}(3)=\underbrace{\int_0^1 f(x_0)dx_0}_{I_0}\underbrace{\int_0^{x_0} \delta (x_1-H(x_0))dx_1}_{I_1} \cdot \nonumber \\ &&\cdot\underbrace{\int_{x_1}^{x_0} \delta(x_2-H^{(2)}(x_0))dx_2}_{I_2} \underbrace{\int_{0}^{x_2} \delta(z_1-H^{(3)}(x_0))dz_1}_{I_3} \underbrace{\int_{x_0}^1 \delta(x_3-H^{(4)}(x_0))dx_3}_{I_4},
\end{eqnarray}
where again each integral $I_m$ contributes with a rescaling of the integration range in $I_0$ according to the solutions of

\begin{eqnarray}
I_1\rightarrow H(x_0)<x_0\nonumber\\
I_2\rightarrow H^{2}(x_0)<x_0\nonumber\\
I_3\rightarrow H^{3}(x_0)<H^{2}(x_0)\nonumber\\
I_4\rightarrow H^{4}(x_0)>x_0
\label{GEN3}
\end{eqnarray}
Proceeding as before, we have:
$$H(x_0)<x_0\Rightarrow x_0 \in [3/4,1].$$
On top of that, the second inequality reduces this range further:
$$H(x_0)<x_0 \cap H^{(2)}(x_0)<x_0 \Rightarrow x_0 \in \bigg[\frac{5+\sqrt{5}}{8},1\bigg],$$
The third inequality reduce it further on:
$$H(x_0)<x_0\  \cap\  H^{(2)}(x_0)<x_0 \ \cap\ H^{(3)}(x_0)< H^{(2)}(x_0) \Rightarrow x_0 \in \bigg[\frac{5+\sqrt{5}}{8},\frac{1}{2}+\frac{\sqrt 3}{4}\bigg]$$
and finally
$$H(x_0)<x_0\  \cap\  H^{(2)}(x_0)<x_0 \ \cap\ H^{(3)}(x_0)< H^{(2)}(x_0) \ H^{4}(x_0)>x_0\Rightarrow x_0 \in [\frac{5+\sqrt{5}}{8}, .925]$$
and therefore the correction of order $\alpha=1$ is
$$P_\rightarrowtail^{(1)}(3)\approx \int_{0.904}^{0.925} \frac{dx_0}{\pi \sqrt{x(1-x)}}=0.024$$
Summarising, up to first order, the predicted probability  
$$P_\rightarrowtail(3)\approx P_\rightarrowtail^{(0)}(3)+P_\rightarrowtail^{(1)}(3)=0.0555$$

\noindent $\underline{\alpha>1}$\\
Before addressing the general high order correction $P_\rightarrowtail^{(\alpha>1)}(3)$, we can say a word about forbidden diagrams. Note that the number of diagrams with $\alpha$ hidden variables can be further labelled using a full binary tree. The seed of the tree is a node that corresponds to the diagram with no hidden variables. This node has a total of $k-1$ offsprings, as the first hidden variable can be placed after one of the $k-2$ variables that are not bounding variables. From there, each node has always two children, labelled $L$ (if the new hidden variable is placed to the left) or $R$ (right). Therefore, any diagram can be uniquely labelled. Diagrams with $\alpha$ hidden variables all of them placed after $x_j$ will be labelled as $j$ followed by a string of letters $L$ or $R$ of size $p-1$.\\
On the other hand, note that all the diagrams that are descendants (offsprings, offsprings of offsprings, etc) of a forbidden diagram are, trivially, also forbidden diagrams. Therefore, it is easy to prove that for $k=3$, at order $\alpha>1$, out of the possible $2^{p}$ diagrams $2^{p-1}$ are forbidden. Moreover, forbidden patterns appear at all levels, in a hierarchical way. Accordingly, only one diagram contributes at order $\alpha$, labelled as $R\underbrace{LL \cdots L}_{\alpha-1}$, whose probability reads
$$P_\rightarrowtail^{(\alpha)}(3)=\int_u^v f(x_0)dx_0,$$
where $[u,v]$ is the range of values of $x_0$ which are solution to the set of inequalities
\begin{eqnarray}
H(x_0)<x_0\nonumber\\
H^{(2)}(x_0)<x_0\nonumber\\
H^{(3)}(x_0)<H^{(2)}(x_0)\nonumber\\
H^{(4)}(x_0)<H^{(2)}(x_0)\nonumber\\
\cdots \nonumber\\
H^{(2+\alpha)}(x_0)<H^{(2)}(x_0)\nonumber\\
H^{(3+\alpha)}(x_0)>x_0
\label{GEN4}
\end{eqnarray}

\noindent Abel-Ruffini's theorem precludes closed solutions for $\alpha\geq 0$, however such corrections can be calculated up to arbitrary precision as they are based on univariate algebraic equations. When we try to calculate some approximants, we find
\begin{eqnarray}
P_\rightarrowtail^{(2)}(3)=P_\rightarrowtail^{(3)}(3)=P_\rightarrowtail^{(4)}(3)=0
\end{eqnarray}
as the intersection range for these cases is empty. By induction we can prove that 
$$P_\rightarrowtail^{(\alpha>1)}(3)=0,$$ and therefore 
the result up to first order is indeed exact:
\begin{eqnarray}
&&P_\rightarrow^{(0)}(3)=0.0315 \nonumber \\
&&P_\rightarrow^{(1)}(3)=0.024 \nonumber \\
&&P_\rightarrow^{(\alpha >1 )}(3)=0\nonumber \\
&&\Rightarrow P_\rightarrow(3)=P_\rightarrow^{(0)}(3)+P_\rightarrow^{(1)}(3)\approx 0.0555
\end{eqnarray}

\noindent In what follows we investigate a general expression for the zeroth order of a general degree $k$. As this has already been calculated for $k\leq3$, let us start with $P_\rightarrowtail(4)$. The zeroth order diagram reads 
\begin{eqnarray}
&&P_\rightarrowtail^{(0)}(4)=\underbrace{\int_0^1 f(x_0)dx_0}_{I_0}\underbrace{\int_0^{x_0} \delta (x_1-H(x_0))dx_1}_{I_1} \cdot \nonumber \\ &&\cdot\underbrace{\int_{x_1}^{x_0} \delta(x_2-H^{(2)}(x_0))dx_2}_{I_2} \underbrace{\int_{x_2}^{x_0} \delta(x_3-H^{(3)}(x_0))dx_3}_{I_3} \underbrace{\int_{x_0}^1 \delta(x_4-H^{(4)}(x_0))dx_4}_{I_4}=\nonumber \\
&&\int_u^v f(x_0)dx_0,
\end{eqnarray}
where $[u,v]$ is the range of values of $x_0$ which are solution to the set of inequalities
\begin{eqnarray}
H(x_0)<x_0\nonumber\\
H^{2}(x_0)<x_0\nonumber\\
H^{3}(x_0)<x_0\nonumber\\
H^{4}(x_0)>x_0
\label{GEN5}
\end{eqnarray}
A pattern is now evident and a formal expression for the zeroth order diagram up to arbitrary degree $k$ can be found by induction:
\begin{eqnarray}
P_\rightarrowtail^{(0)}(k)= \int_u^v f(x_0)dx_0, \ 
\end{eqnarray}
 where $[u,v]$ is the range of values of $x_0$ which are solution to the system of inequalities 
\begin{eqnarray}
&&H^{i}(x_0)<x_0, i=1,\dots,k-1 \nonumber \\
&&H^{k}(x_0)>x_0
\label{GEN6}
\end{eqnarray}
These inequalities can be solved up to arbitrary numerical precision. Some results include:
\begin{eqnarray}
P_\rightarrowtail^{(0)}(4)=0.03 \nonumber \\
P_\rightarrowtail^{(0)}(5)= 0.02\nonumber \\
P_\rightarrowtail^{(0)}(6)= 0.014\nonumber 
\end{eqnarray}

\noindent Note that the analysis performed here is far more general than the case of the logistic map, and indeed the set of inequalities (\ref{GEN1}, \ref{GEN2}, \ref{GEN3}, \ref{GEN4}, \ref{GEN5}, \ref{GEN6}) holds in general provided the map has an invariant, $L_1$-integrable $f(x)$. To evaluate the accuracy and convergence speed of the perturbative expansion in the concrete case of the fully chaotic logistic map, we make some numerical simulations. We generate time series of $2^{20}$ data extracted from a logistic map with the same characteristics, transform the series into a DHVg and compute its degree distribution. A comparison with our theory is shown in table \ref{table1}.
The first three terms $k=1,2,3$ show a perfect agreement with numerics because in those cases the theory was exact. For the rest, we can see that the free field solution (zeroth order) gets more accurate for larger values of $k$.

\begin{table}
\begin{ruledtabular}
\begin{tabular}{cccc}
degree $k$ &$P_\rightarrowtail(k)$ (numerics)&$P_\rightarrowtail^{(0)}(k)$  &$P_\rightarrowtail^{(0)}(k)+P_\rightarrowtail^{(1)}(k)$\\
\hline
1&0.666&0.666&0.666\\
2&0.133&$\frac{2}{15}=0.133$&$\frac{2}{15}$\\
3&0.055&0.0315&0.055\\
4&0.044&0.03&\\
5&0.03&0.02&\\
6&0.02&0.014&\\
\end{tabular}
\end{ruledtabular}
\caption{\label{table1}Comparison of numerical results and theoretical predictions for the out degree distribution of DHVG associated to the fully chaotic logistic map
$x_{t+1}=4x_t(1-x_t)$. Numerics are the result of simulations, where a trajectory of $2^{20}$ time steps extracted from the map is mapped to its DHVG and the degree distribution is numerically computed. For $k=1,2$ perturbation theory at zeroth order is exact due to the restriction of forbidden patterns (see the text). For $k=3$ we need to go up to first order to reach an accurate result.}
\end{table}

\subsection{Quasiperiodic dynamics}
\begin{figure}
\centering
\includegraphics[width=0.9\columnwidth]{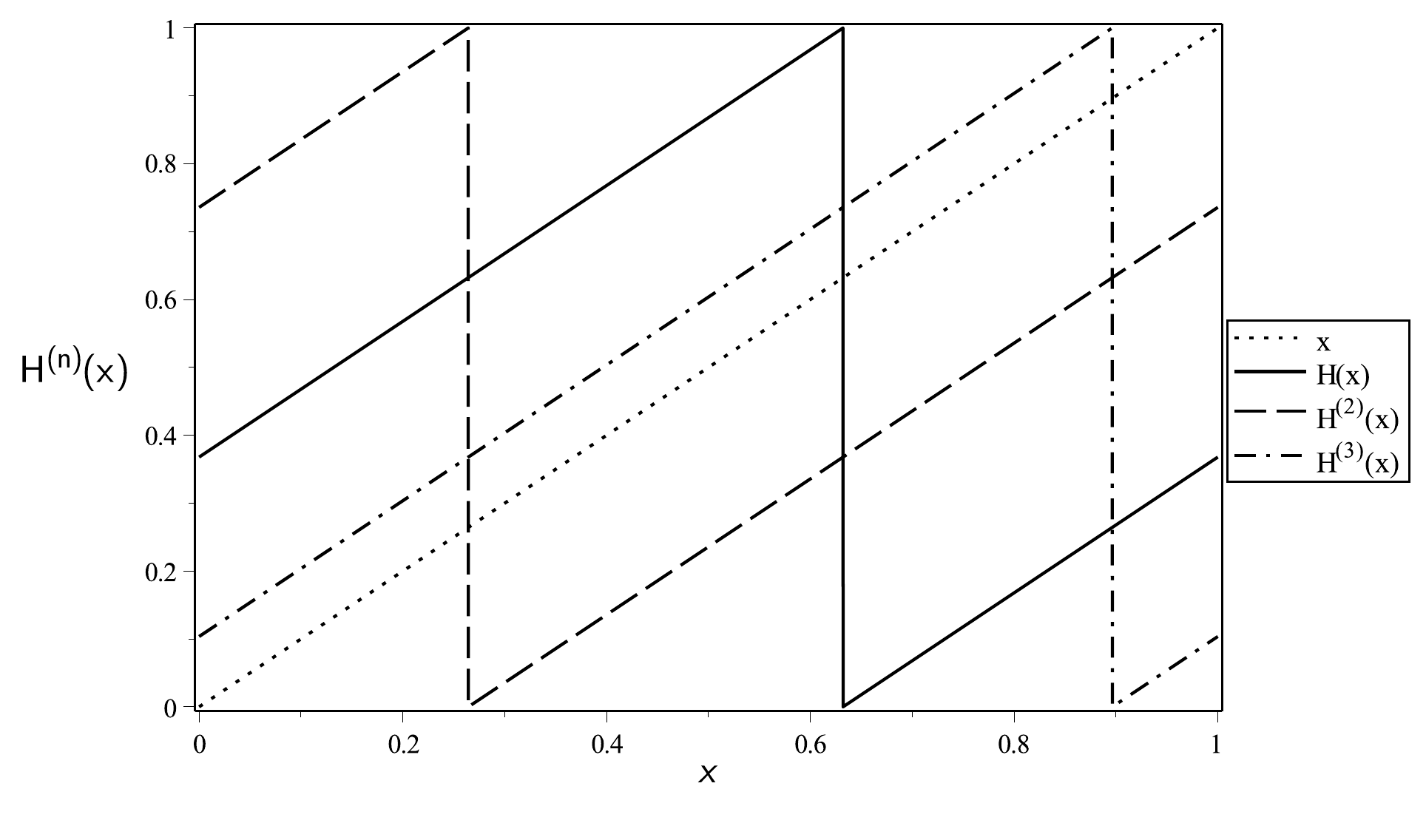}
\caption{Cobweb plot of the quasiperiodic map $H(x)=x +\omega \mod 1$, where $\omega$ is irrational (in this figure $\omega=\exp(-1)$) and some of its iterates $H^{(n)}(x)$. The specific intervals where $H^{(n)}(x)>(<)x$.}
\label{qp}
\end{figure}

We now address a paradigmatic example of a nonchaotic deterministic dynamics which is nonetheless aperiodic. Consider the Poincare map associated with a flow having two incommensurate frequencies \cite{just}, or simply an irrational rotation
$$H(x)=(x+\omega) \mod 1, $$
where $\omega$ is irrational which without loss of generality we may assume to be smaller than one (otherwise, the map is equivalent to $H(x)=(x+\omega-\lfloor \omega \rfloor) \mod 1 $.) Trajectories are aperiodic and fill up densely the interval $[0,1]$. The map is ergodic and has a uniform invariant measure $U[0,1]$, albeit having zero entropy. In our notation,
$$f(x)=1,$$
therefore the integration rescaling $[a,b]\rightarrow [u,v]$ exposed in the previous section yields a general diagram whose probability correction is simply $v-u$. A cobweb diagram of some iterations of this map is shown in figure \ref{qp}.
The nth-iteration of the map can be simply written as
$$H^{(n)}(x)=(x+n\omega) \mod 1,$$
what yields a simpler set of criteria:
\begin{eqnarray}
H^{(n)}(x)<x &\Longleftrightarrow & x > 1- n\omega\mod 1,\ \forall n \geq 1\nonumber \\
&\Longleftrightarrow & \left\{
\begin{array}{l}
x>1-n\omega, \   n<1/\omega, \\
x>1-n\omega + \lfloor n\omega \rfloor,  n>1/\omega  %
\end{array}%
\right. 
\end{eqnarray}

The calculations are conceptually equivalent to the ones in previous section, although the concrete techniques differ as this map does not has fixed points. In what follows we show a summary of the first terms. First,
\begin{equation}
P_\rightarrowtail(1)=\int_0^1 dx_0\int_{x_0}^1 \delta (x_1-H(x_0))dx_1.
\end{equation}
The second integral is $1$ for $H(x_0)>x_0$ (and zero otherwise), that is, for $x_0<1-\omega$. Therefore,
\begin{equation}
P_\rightarrowtail(1)=\int_0^{1-\omega} dx_0=1-\omega
\end{equation}
For $P_\rightarrowtail(2)$ we expand again in a perturbation series
$$P_\rightarrowtail(2) = \sum_{\alpha=0}^{\infty}P_\rightarrowtail^{(\alpha)}(2),$$
whose free field solution $\alpha=0$ reduces to the intersection
$$[1-\omega, 1] \cap [0,1-2\omega]$$
This interval is non null for $\omega>1/2$ and doesn't contribute otherwise (forbidden pattern). For $\omega>1/2$ its contribution is
$$P_\rightarrowtail^{(0)}(2)=\lfloor 2\omega \rfloor - \omega$$
Similarly, the correction of order $\alpha=1$ reduces to the intersection
\begin{eqnarray}
(H(x)< x) \cap (H^{(2)}(x)< H(x)) \cap (H^{(3)}(x)>x) 
\end{eqnarray}
which is null for $\omega >1/2$. By iteration, it can be proved that 
$$P_\rightarrowtail(2) = \sum_{\alpha=1}^{\infty}P_\rightarrowtail^{(\alpha)}(2)=0, \  \forall \omega$$
Similarly, the free field solution for $P_\rightarrowtail(3)$ reduces to
\begin{eqnarray}
(H(x)< x) \cap (H^{(2)}(x)< x) \cap (H^{(3)}(x)>x). 
\end{eqnarray}
The first inequality requires $$x \in [1-\omega, 1], \ \forall \omega.$$ The second one requires
$$x  \left\{
\begin{array}{l}
\in [1-2\omega, 1], \ \forall \omega < 1/2, \\
\in [1-2\omega + \lfloor 2\omega \rfloor,1], \ \forall \omega > 1/2  %
\end{array}%
\right. 
 $$
 and the third requires
 $$x  \left\{
\begin{array}{l}
\in [0, 1-3\omega], \ \forall \omega < 1/3, \\
\in [0,1-3\omega+\lfloor 3\omega \rfloor], \ \forall \omega > 1/3  %
\end{array}%
\right. 
 $$
So depending on the specific value of $\omega$, this contribution can be a finite or null value. In table \ref{qp} we summarise these theoretical estimations  for different values of $\omega$, along with the results of numerical simulations.

\begin{table}
\begin{ruledtabular}
\begin{tabular}{ccccc}
 $k$ &$\omega$& $P_\rightarrowtail^{(0)}(k)$ & Numerics\\
\hline
1&$\exp(-1)$&0.632&0.632\\
2&$\exp(-1)$&0&0\\
3&$\exp(-1)$&0.264&0.264\\
1&$\pi-3$&0.858&0.858\\
2&$\pi-3$&0&0\\
3&$\pi-3$&0&0\\

1&$\phi^{-1}$&0.382&0.382\\
2&$\phi^{-1}$&0.382&0.382\\
3&$\phi^{-1}$&0&0.146\\

1&$\sqrt 2-1$&0.5857&0.5857\\
2&$\sqrt 2-1$&0&0\\
3&$\sqrt 2-1$&0.171&0.3431\\

\end{tabular}
\end{ruledtabular}
\caption{\label{tableqp}Comparison of numerical results and theoretical predictions for the out degree distribution of DHVG associated to the irrational rotation
$x_{t+1}=x_t + \omega \mod 1$, where $\omega$ is an irrational number $\in (0,1)$. Numerics are the result of simulations, where a trajectory of $2^{20}$ time steps extracted from the map is transformed into a DHVG and the degree distribution is numerically computed. For $k=1,2$ perturbation theory at zeroth order is exact due to the restriction of forbidden patterns (see the text). For $k=3$ the zeroth order is an accurate approximation of the experimental solution for some values of $\omega$.}
\end{table}

To summarise, in the last sections we have developed a formal theory to compute concrete degree probabilities $P_\rightarrow(k)$ of DHVGs associated to different types of Markovian dynamics. In the next section we take a totally different approach and investigate how to make use of calculus of variations to calculate the whole degree distribution associated to these processes, and for completeness we extend these to both DHVG and HVGs.

\section{Variational techniques}

\noindent Let the Shannon entropy over the degree distribution of an HVG, ${\Phi }\{P(k)\}$ be
$${\Phi}\{P(k)\}=-\sum_{k=2}^{\infty} P(k) \log P(k)$$
On the other hand, periodic series of period $T$ yield HVGs with mean degree \cite{chaos, Gutin}
$$\langle k \rangle_{\text{HVG}}=4\bigg(1-\frac{1}{2T}\bigg),$$
which means that aperiodic series reach the upper bound $\langle k \rangle_{\text{HVG}}=4$, independently of whether the underlying process is deterministic or stochastic. In previous research \cite{chaos} it was shown that  HVGs associated to uncorrelated random processes are maximally entropic, in the sense that the degree distributions that maximise $S$, if we require $P(k)$ to be normalised and that $\langle k \rangle_{\text{HVG}}=4$, reduces to equation \ref{uncorr}. In other words, uncorrelated random processes are maximally ${\Phi}_{\text{HVG}}$-entropic processes.\\

\noindent In this section we first show that a similar principle holds for DHVGs and that the DHVGs associated to uncorrelated random processes are maximally entropic.
We then conjecture that stochastic and chaotic processes are also maximally entropic restricted to further requirements: whereas they are also aperiodic 
and the mean degree is maximal, they have correlations recorded by $P_\rightarrow(k)$ for small values of the degree. To test this conjecture,
we formulate a similar MaxEnt problem where the values of $P_\rightarrow(s), s=1,2,3$ (calculated exactly in previous sections) are fixed. We also explore this principle for HVGs.

\subsection{Uncorrelated random processes: maximally entropic DHVGs}

First of all note a periodic series of period $T$ (generated from a stochastic or deterministic Markovian dynamical system) maps to a DHVG whose mean in and out degree read
$$\langle k_\leftarrow \rangle_{\text{DHVG}}=\langle k_\rightarrow \rangle_{\text{DHVG}}=2\bigg(1-\frac{1}{2T}\bigg)$$
(the proof trivially follows from the one for the HVG \cite{chaos} and is therefore skipped in this presentation). The maximal mean degree of a DHVG is therefore $\langle k_\leftarrow \rangle_{\text{DHVG}}=\langle k_\rightarrow \rangle_{\text{DHVG}}=2$, reached for aperiodic series.\\
  
\noindent In order to calculate the degree distribution $P_\rightarrow(k)$ that maximises the entropy of a DHVG ${\Phi}\{P_\rightarrow(k)\}=\sum_{k=1}^{\infty} P_\rightarrow(k) \log P_\rightarrow(k)$, let us define the following Lagrangian
\begin{eqnarray}
{\cal L}_\rightarrow= -\sum_{k=1}^{\infty} P_\rightarrow(k) \log P_\rightarrow(k) -\lambda_0\bigg(\sum_{k=1}^{\infty} P_\rightarrow(k) -1\bigg)- \lambda_1\bigg(\sum_{k=1}^{\infty} k P_\rightarrow(k) - 2\bigg)
\end{eqnarray}
The extremum condition reads
\begin{eqnarray}
\frac{\partial \cal L_\rightarrow}{\partial P_\rightarrow(k)}=0 \nonumber
\end{eqnarray}
whose general solution is an exponential function
$$P_\rightarrow(k)= \exp(-\lambda_0-\lambda_1 k),$$
where $\lambda_0$ and $\lambda_1$ are Lagrange multipliers that can be solved from the constraints. The first constraint is the distribution normalisation
\begin{equation}
\sum_{k=1}^{\infty }{e^{-\lambda _{0}-\lambda _{1}k}}=1,
\label{RR0}
\end{equation}%
which implies the following relation between $\lambda _{0}$ and $\lambda _{1}$ 
\begin{equation}
e^{\lambda _{0}}=\frac{1}{e^{\lambda_1 - 1}},
\label{RR1}
\end{equation}%
using the sum of a trigonometric series. Now, the second restriction is on the mean degree:
\begin{equation}
\sum_{k=1}^{\infty} ke^{-\lambda_{0}-\lambda_{1}k} =e^{-\lambda_{0}}\sum_{k=1}^{\infty }ke^{-\lambda_{1}k}=2
\label{RR2}
\end{equation}%
Now, notice that 
$$ \frac{\partial}{\partial \lambda_1} \bigg( \sum e^{-\lambda_1 k} \bigg)= - \sum k e^{-\lambda_1 k}$$
and therefore differentiating equation \ref{RR0} with \ref{RR1} gives
$$  \sum k e^{-\lambda_1 k}=\frac{e^{\lambda_1}}{(e^{\lambda_1}-1)^2}
$$
which we introduce in equation \ref{RR2} to find 
$$\frac{e^{\lambda_1}}{(e^{\lambda_1}-1)^2}=2e^{\lambda_0}.$$
This last expression, together with equation \ref{RR1}, solves $\lambda_0$ and $\lambda_1$
$$\lambda_0=0 ,\  \lambda_1=\log 2.$$
Therefore, the degree distribution that maximizes ${\cal L}_\rightarrow$ is 
\begin{equation*}
P(k)=\bigg(\frac{1}{2}\bigg)^k, \ k=1,2,...
\end{equation*}%
which is the result found for uncorrelated random processes (theorem 2). We conclude that the DHVG of these processes are maximally entropic or, equivalently, that uncorrelated random processes are maximally $\Phi_{\text{DHVG}}$-entropic.

\subsection{Ornstein-Uhlenbeck and logistic map I: DHVGS} 
\noindent There is numerical evidence \cite{irrev} that suggests that both chaotic and some correlated stochastic processes also have DHVGs with exponentially decaying degree distributions, for $k$ sufficiently large. An heuristic justificaton argues that for long times, chaotic and random uncorrelated processes cannot be distinguished, and the same holds for correlated stochastic processes with a fast decaying correlation function (Ornstein-Uhlenbeck). Therefore we expect that the degree distribution of these processes deviate from the i.i.d. theory for short times, whereas we conjecture that a similar entropic extremization may take place for sufficiently large values. The question of course is to determine what "sufficiently large" means, that is, which is the minimal value of $k$ where the exponential decay starts to be a good approximation. We define an $s$- dependent Lagrangian ${\cal L}_\rightarrow(s)$, such that
\begin{eqnarray}
{\cal L}_\rightarrow(1) &=& -\sum_{k=2}^{\infty} Q_\rightarrow(k) \log Q_\rightarrow(k) -\lambda_0\bigg(\sum_{k=2}^{\infty} Q_\rightarrow(k) -(1-p_1)\bigg)- \nonumber\\
&&- \lambda_1\bigg(\sum_{k=2}^{\infty} kQ_\rightarrow(k) + p_1-2\bigg),
\end{eqnarray}
where $p_1=P_\rightarrow(1)$,
\begin{eqnarray}
{\cal L}_\rightarrow(2) &=& -\sum_{k=3}^{\infty} Q_\rightarrow(k) \log Q_\rightarrow(k) -\lambda_0\bigg(\sum_{k=3}^{\infty} Q_\rightarrow(k) -(1-p_1-p_2)\bigg)- \nonumber\\
&&- \lambda_1\bigg(\sum_{k=3}^{\infty} kQ_\rightarrow(k) + p_1+2p_2-2\bigg),
\end{eqnarray}
where $p_2=P_\rightarrow(2)$, and
\begin{eqnarray}
{\cal L}_\rightarrow(3) &=& -\sum_{k=4}^{\infty} Q_\rightarrow(k) \log Q_\rightarrow(k) -\lambda_0\bigg(\sum_{k=4}^{\infty} Q_\rightarrow(k) -(1-p_1-p_2-p_3)\bigg)- \nonumber\\
&&- \lambda_1\bigg(\sum_{k=4}^{\infty} kQ_\rightarrow(k) + p_1+2p_2+3p_3-2\bigg).
\end{eqnarray}
where $p_3=P_\rightarrow(3)$. From previous sections, we have learned how to compute these terms, either in an exact or in a perturbative way. In this section we only use exact results for these terms. Our approach consists in finding $Q_\rightarrow(k)$ such that
\begin{equation}
P_\rightarrow(k)=\left\{
\begin{array}{l}
p_k, \   k\leq s, \\
Q_\rightarrow(k), \  k > s.%
\end{array}%
\right.   \label{pkRG}
\end{equation}%
If these graphs are still maximally entropic, then a MaxEnt argument should predict the correct shape of the full distribution.\\
After a little algebra we come to a system of two equations for the Lagrange multipliers $\{\lambda_0, \lambda_1\}_s$, such that for $s=1$ we have
\begin{eqnarray}
e^{\lambda_0}=\frac{1}{1-p_1} \bigg(\frac{1}{e^{\lambda_1}-1}-e^{-\lambda_1}\bigg)\nonumber \\
e^{\lambda_0}=\frac{1}{2-p_1}\bigg( \frac{e^{\lambda_1}}{(e^{\lambda_1}-1)^2 } - e^{-\lambda_1}  \bigg),
\end{eqnarray}
for $s=2$, $\{\lambda_0, \lambda_1\}_2$ fulfills
\begin{eqnarray}
e^{\lambda_0}=\frac{1}{1-p_1-p_2} \bigg(\frac{1}{e^{\lambda_1}-1}-e^{-\lambda_1}-e^{-2\lambda_1}\bigg)\nonumber \\
e^{\lambda_0}=\frac{1}{2-p_1-2p_2}\bigg( \frac{e^{\lambda_1}}{(e^{\lambda_1}-1)^2 } - e^{-\lambda_1} -2e^{-2\lambda_1} \bigg)
\end{eqnarray}
while for $s=3$, $\{\lambda_0, \lambda_1\}_3$ fulfills
\begin{eqnarray}
e^{\lambda_0}=\frac{1}{1-p_1-p_2-p_3} \bigg(\frac{1}{e^{\lambda_1}-1}-e^{-\lambda_1}-e^{-2\lambda_1}e^{-3\lambda_1}\bigg)\nonumber \\
e^{\lambda_0}=\frac{1}{2-p_1-2p_2-3p_3}\bigg( \frac{e^{\lambda_1}}{(e^{\lambda_1}-1)^2 } - e^{-\lambda_1} -2e^{-2\lambda_1} -3e^{-3\lambda_1} \bigg)
\end{eqnarray}
Note again that this formulation can be easily extended for an arbitrary $s$, if needed. Here we will investigate the results for $s=2,3$.\\
\begin{figure}
\centering
\includegraphics[width=0.9\columnwidth]{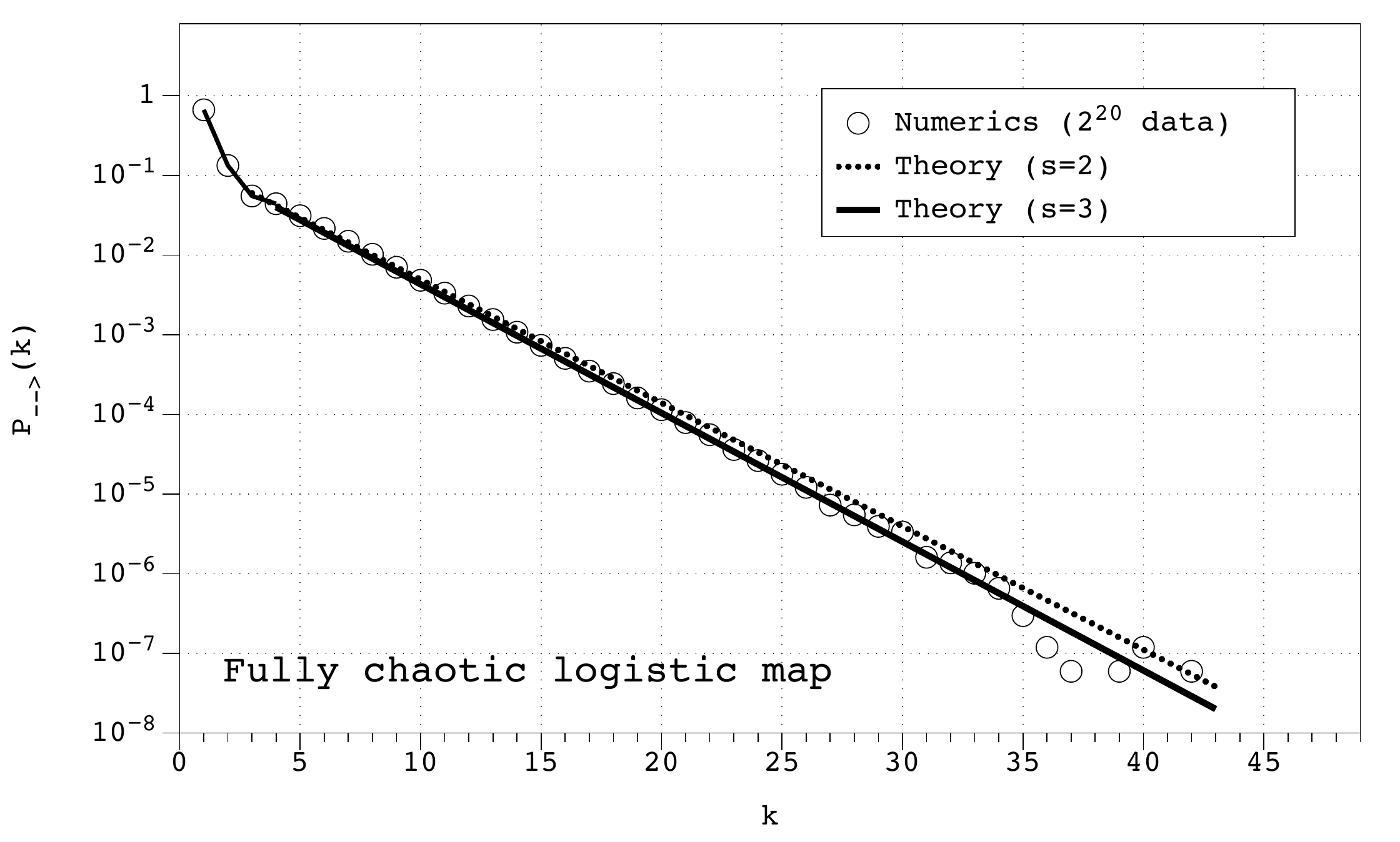}
\caption{Semi log plot of the degree distribution $P_\rightarrow(k)$ of a DHVG associated to a fully chaotic logistic map. Dots are the result of numerical simulations on a time series of $2^{20}$ data, solid lines are the prediction of the variational approach.}
\label{maxent1}
\end{figure}
\noindent Let us start with the fully chaotic logistic map. For ${\cal L}_\rightarrow(2)$, where $p_1=1/3, \ p_2=2/15$, we find
\begin{eqnarray}
&&\lambda_1= \log 2+ \log 5- \log 7 \approx 0.356674944 \nonumber \\
&&\lambda_0= \log \bigg( \frac{1}{1-p_1-p_2} \bigg(\frac{1}{e^{\lambda_1}-1}-e^{-\lambda_1}-e^{-2\lambda_1}\bigg) \bigg)\approx 1.743385883\nonumber
\end{eqnarray}
while for ${\cal L}_\rightarrow(3)$ (where $p_1=1/3, \ p_2=2/15, \ p_3=0.55$), the Lagrange multipliers read $\lambda_1\approx 0.372$, $\lambda_0\approx 1.727$.
In figure  \ref{maxent1} we plot the theoretical prediction of $P_\rightarrow(k)$ along with the numerical simulations obtained for a time series of $2^{20}$ data. The agreement is very good for $s=2$ and even better for $s=3$, what confirms that (i) the chaotic process is also maximally ${\Phi}_{\text{DHVG}}$-entropic up to short range restrictions, (ii) short term correlations are captured in $P_\rightarrow(1), P_\rightarrow(2), P_\rightarrow(3)$, and (ii) that a very good theoretical approximation for its full degree distribution is

\begin{equation}
P_\rightarrow(k)=\left\{
\begin{array}{l}
2/3, \   k=1, \\
2/15, \ k=2, \\
0.55, \ k=3, \\
\exp(0.727-0.372 k), \  k \geq 4.%
\end{array}%
\right.   \label{pkcaos}
\end{equation}%

\begin{figure}
\centering
\includegraphics[width=0.9\columnwidth]{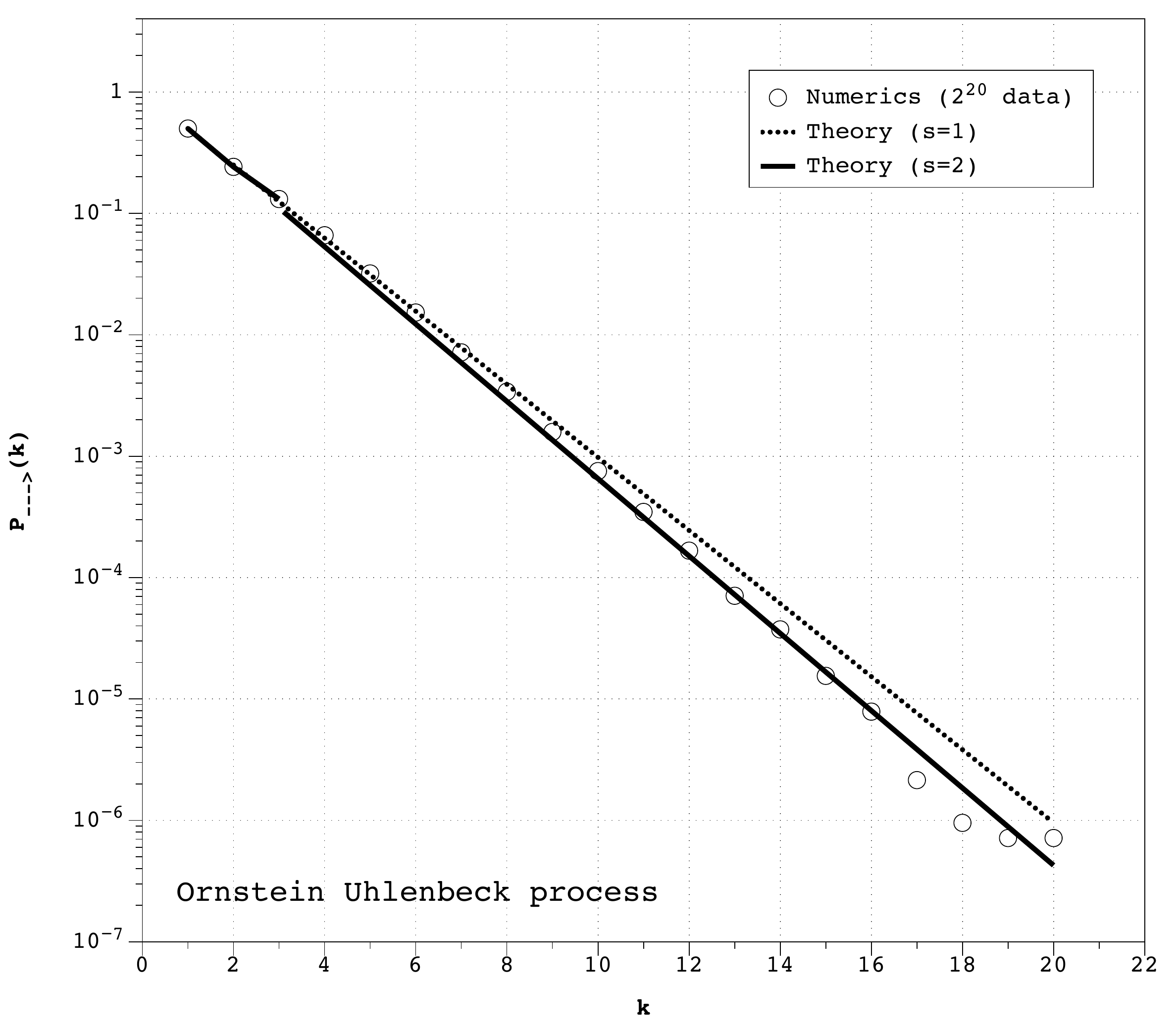}
\caption{Semi log plot of the degree distribution $P_\rightarrow(k)$ of a DHVG associated to an Ornstein-Uhlenbeck with correlation time $\tau=1.0$. Dots are the result of numerical simulations on a time series of $2^{20}$ data, solid lines are the prediction of the variational approach.}
\label{maxent2}
\end{figure}

\noindent For the Ornstein-Uhlenbeck process we only have exact results for $P_\rightarrow(1)$ and $P_\rightarrow(2)$. For ${\cal L}_\rightarrow(1)$ (where $p_1=1/2$), the Lagrange multipliers are $\lambda_1= \log(2)$, $\lambda_0=0$, that is, the prediction for an uncorrelated process. The prediction clearly improves for ${\cal L}_\rightarrow(2)$, as can be seen in figure  \ref{maxent2}, where we plot the theoretical results for $P_\rightarrow(k)$ along with the numerical simulations of the process (series of $2^{20}$ data). This confirms that the Ornstein-Uhlenbeck process is also maximally ${\Phi}_{\text{DHVG}}$-entropic up to short range restrictions. In this case, short term correlations are well captured by $P_\rightarrow(1), P_\rightarrow(2)$. The theoretical degree distribution is in this case
\begin{equation}
P_\rightarrow(k)=\left\{
\begin{array}{l}
1/2, \   k=1, \\
0.24, \ k=2, \\
\exp(-0.733 k), \  k \geq 3.%
\end{array}%
\right.   \label{pkcaos}
\end{equation}%

\subsection{Ornstein-Uhlenbeck and logistic map II: HVGS}
\begin{figure}
\centering
\includegraphics[width=0.9\columnwidth]{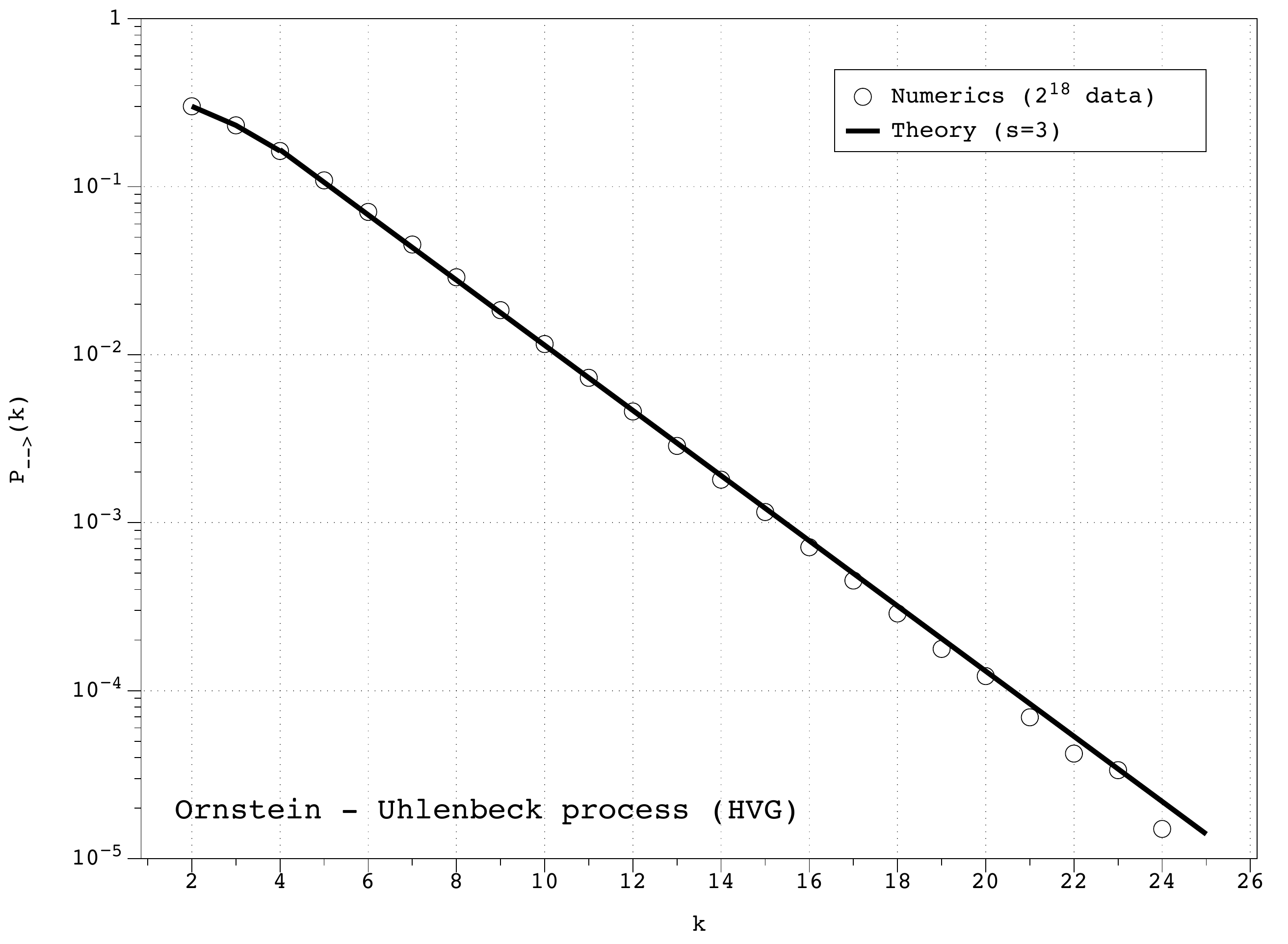}
\caption{Semi log plot of the degree distribution $P(k)$ of a HVG associated to an Ornstein-Uhlenbeck with correlation time $\tau=1.0$. Dots are the result of numerical simulations on a time series of $2^{18}$ data, solid lines are the prediction of the variational approach.}
\label{maxent3}
\end{figure}

Previous numerical evidence suggests that these processes also have HVGs with exponentially distributed degree sequences $P(k)\sim \exp(-\lambda k)$, where $\lambda$ distinguish the purely chaotic from the stochastic ones \cite{toral}. In this last section we repeat the same variational arguments for HVGs associated to both Ornstein-Uhlenbeck and fully chaotic logistic map, and find that an entropy maximization principle predicts accurately the degree distribution. In both cases, the Lagrangian reads

\begin{eqnarray}
{\cal L}(s) &=& -\sum_{k=s+1}^{\infty} Q(k) \log Q(k) -\lambda_0\bigg(\sum_{k=s+1}^{\infty} Q(k) -(1-\sum_{i=2}^sp_i)\bigg)- \nonumber\\
&&- \lambda_1\bigg(\sum_{k=s+1}^{\infty} kQ(k) + \sum_{i=2}^sp_i-4\bigg),
\end{eqnarray}
For the logistic map, $P(2)=P(3)=1/3$ \cite{toral}. Accordingly, a MaxEnt principle up to $s=3$ predicts an exponential decay with $\lambda_1=\log(4/3) \approx 0.28$, to be compared with the numerical estimate $\lambda_1^{\text{num}} \approx 0.26$ \cite{toral}. We expect more accurate predictions of the slope of the distribution for higher values of $s$.\\
For the Ornstein-Uhlenbeck with $\tau=1.0$, analytical estimates of the HVG are $P(2)\approx0.3012, \ P(3)\approx 0.23$ \cite{toral}, which yield under a MaxEnt principle an exponentially decaying degree distribution with $\lambda_1 \approx 0.4467$, to be compared with numerics in figure \ref{maxent3}, finding an excellent match between theory and numerical experiment.\\

\noindent Finally, let us observe that the irrational rotation does not fulfill a MaxEnt principle in this context, as expected (correlations do not decay fast for quasiperiodic dynamics).
 
\section{Discussion}
In this work we have addressed the degree distribution of horizontal visibility graphs (both undirected and directed). These are classes of graphs constructed from the mapping of dynamics via the so called horizontal visibility algorithm, which are currently widely used for nonlinear time series analysis. In the first part we have developed a diagrammatic theory to analytically compute each component of the out degree distribution. In the case of deterministic dynamical systems, we have found that diagrammatic expansions converge fast due to the presence of forbidden diagrams. A more detailed analysis of these diagrams on relation to ordinal patterns \cite{forbidden} is left for future work.\\
Note that the use of diagrammatic expansions in nonlinear dynamics is an idea first advanced by Percival and Vivaldi \cite{vivaldi}, and later explored by Beck \cite{beck}. Our theory works for dynamical systems that have a $L_1$-integrable invariant measure. Accordingly, dissipative chaotic maps with strange attractors are left out: in those cases, a general measure theory over fractal sets should be followed, this is also an open problem for further research.\\
The theory also applies to stochastic processes with the Markovian property. In this work we dealt with Guassian processes, whose high order diagrammatic corrections are, from a practical point of view, computationally expensive. Further work could explore the saddle-point method to approximate these higher order corrections, to speed up the calculations.\\
In a second part, we have used Jayne's MaxEnt variational approach to obtain estimates of the full degree distributions associated to the chaotic and stochastic cases, finding that these are maximally entropic correlated processes.\\

\noindent To conclude, note that the theory developed in this work can be further extended to other types of Markovian dynamics. In this sense, previous numerical evidence on the degree distribution associated to different types of dynamics
could be analytically addressed via this methodology.

\bibliography{apssamp}

\end{document}